\documentclass[11pt]{article}

\usepackage[T1]{fontenc}
\usepackage[utf8]{inputenc}
\usepackage{lmodern}
\usepackage[margin=1in]{geometry}
\usepackage{amsthm,amsmath,amsfonts,amssymb}
\usepackage[authoryear]{natbib}
\usepackage[colorlinks,citecolor=blue,urlcolor=blue]{hyperref}
\usepackage{graphicx}
\usepackage{algorithm}
\usepackage{algpseudocode}
\usepackage{booktabs}
\usepackage{makecell}
\usepackage{enumitem}
\usepackage{bm}
\usepackage{xcolor}
\newcommand{\mathbbm}[1]{\mathbf{#1}}
\usepackage{float}
\usepackage{multirow}

\newtheorem{lemma}{Lemma}
\newtheorem{theorem}{Theorem}
\newtheorem{proposition}{Proposition}
\newtheorem{assumption}{Assumption}

\newcommand{\ind}{\perp\!\!\!\!\perp}

\title{Double machine learning to estimate the effects of multiple treatments and their interactions}
\author{%
\normalsize
Qingyan Xiang\textsuperscript{1,*}, Yubai Yuan\textsuperscript{2}, Dongyuan Song\textsuperscript{3},\\
Usman J.~Wudil\textsuperscript{4}, Muktar H.~Aliyu\textsuperscript{4}, C.~William Wester\textsuperscript{4},\\
Bryan E.~Shepherd\textsuperscript{1}\\[0.75em]
\small \textsuperscript{1}Department of Biostatistics, Vanderbilt University Medical Center\\
\small \textsuperscript{2}Department of Statistics, Penn State University\\
\small \textsuperscript{3}Department of Genetics and Genome Sciences, University of Connecticut Health Center\\
\small \textsuperscript{4}Vanderbilt Institute for Global Health, Vanderbilt University Medical Center\\
\small *Corresponding author: \texttt{qingyan.xiang@vumc.org}}
\date{}

\begin{document}

\maketitle

\begin{abstract}
Causal inference literature has extensively focused on binary treatments, with increasing attention to multi-valued treatments. However, methods for multiple simultaneously assigned treatments are still understudied, despite their practical importance. This paper introduces two settings: (1) estimating the effects of multiple concurrent treatments of different types (binary, categorical, and continuous) and the effects of treatment interactions, and (2) estimating the average treatment effect across categories of multi-valued regimens. To obtain robust estimates for both settings, we study a class of methods based on the double machine learning framework. By using machine learning to flexibly model confounding relationships, these methods are well suited to complex settings with multiple treatments or regimens. Bias and overfitting arising from machine learning models can be overcome through Neyman orthogonality and cross-fitting. We further establish the asymptotic distribution of our estimators and derive variance estimators for statistical inference. Simulations demonstrate the performance of our methods. Finally, we apply the methods to study the effects of three treatments on HIV-associated kidney disease in an adult HIV cohort of 2455 participants in Nigeria.
\end{abstract}


%%%%%%%%%%%%%%%%%%%%%%%%%%%%%%%%%%%%%%%%%%%%%%
%% Please use \tableofcontents for articles %%
%% with 50 pages and more                   %%
%%%%%%%%%%%%%%%%%%%%%%%%%%%%%%%%%%%%%%%%%%%%%%
%\tableofcontents

%%%%%%%%%%%%%%%%%%%%%%%%%%%%%%%%%%%%%%%%%%%%%%
%%%% Main text entry area:
\section{Introduction}
\label{sec: intro}
\subsection{Background and motivation}
The development of causal inference methods has made significant progress to address confounding issues in  observational studies \citep{hernan2024causal, ding2024first}. However, many causal inference methods focus on a binary treatment, comparing the effect of a treatment versus a control \citep{rosenbaum1983central, gutman2015estimation}. These methods can be inadequate for more complex settings involving multiple treatments, especially when treatment interactions are present. Such settings are not only methodologically challenging but also practically important. For example, in an HIV study, researchers seek to evaluate the side effect of multiple treatments and their interactions on participants' kidney function. To draw meaningful conclusions, it is essential to develop robust causal inference methods that can handle the settings of multiple treatments and their interactions.

Besides the lack of proper methods for multiple treatments, analyzing observational data presents additional challenges, including complex confounding relationships (e.g. non-linear patterns and higher-order interactions)  and high-dimensional confounding covariates \citep{huber2013performance, wan2018evaluation, berisha2021digital}. These challenges complicate the estimation process and increase the potential for bias \citep{d2021overlap, johnstone2009statistical};  the complexity increases further with multiple treatments, as the number of variables and interactions grows.  Given these challenges, machine learning methods become a natural choice because of their strengths in modeling high-dimensional data, non-linear patterns, and interactions \citep{jiang2020supervised}. However,  direct use of machine learning in causal inference is not a ``free lunch''. As machine learning methods specialize in prediction instead of estimation, they are subject to overfitting and regularization bias \citep{neyshabur2014search, mehrabi2021survey,kernbach2022foundations}, which can further cause bias in estimating treatment effects.

To address these challenges, we study a class of methods for multiple treatments based on double machine learning (DML) \citep{chernozhukov2018double}, a framework that uses machine learning methods to estimate causal effects while overcoming the potential bias induced by regularization and overfitting. Specifically, we contribute to the literature in two broadly applicable settings: (1) a DML partial linear model to estimate the effects of multiple concurrent treatments of different types (binary, categorical, and continuous), as well as their interactions, and (2) a DML interactive model to estimate the average treatment effect (ATE) across categories of multi-valued regimens. For both settings, we define estimands, propose estimators, and develop corresponding algorithms for effect estimation. We further describe the asymptotic normality of the proposed estimators and thereby derive the analytical variance estimator for inference.

\subsection{The northern Nigeria HIV cohort}

We apply our method to study the effect of three different treatments (and their interactions) on chronic kidney disease in an HIV cohort of 2455 participants in northern Nigeria. This application is motivated by the rising incidence of HIV-associated kidney disease \citep{wudil2021apolipoprotein}. Since people with HIV are typically on multiple treatments, studying the side effects of these treatments on kidney function can improve our understanding of HIV-associated kidney disease.

In this study, kidney function is assessed using the estimated glomerular filtration rate (eGFR), a measure of the severity of chronic kidney disease. Besides the three treatments, the study included a total of 15 demographic and clinical variables, which may exhibit complex relationships with treatment assignment and kidney function. This application demonstrates how our method addresses challenges arising from multiple treatments and complex confounding structures.

\subsection{Existing work: multi-valued treatments, or multiple concurrent treatments}
Recent years have seen a growth in causal inference methods for multi-valued treatments. \citet{imbens2000role} and \citet{imai2004causal} extended the work of \citet{rosenbaum1983central} by proposing the generalized propensity score for multi-valued treatments. Thereafter, many methods have been adapted for multi-valued treatments using the generalized propensity scores: inverse propensity score weighting (IPW) \citep{feng2012generalized, mccaffrey2013tutorial}, propensity score adjustment \citep{linden2016estimating}, and propensity score matching (PSM) \citep{yang2016propensity, lopez2017estimation}. Other methods have also been extended for multi-valued treatments, including balancing weights \citep{li2019propensity}, Bayes additive regression trees \citep{hu2020estimation, hu2022flexible}, and targeted maximum likelihood estimation \citep{rose2019double}.

The aforementioned literature focuses on ``multi-valued treatments,'' where each subject receives a single level of treatment that is modeled as a categorical variable $A_i=k$ for $k\in \{1,\ldots{},K\}$. Therefore, the generalized propensity score is estimated for each category, and the focus is typically on the pairwise contrast, such as ATE between treatment categories. However, in many fields---such as clinical \citep{wudil2021apolipoprotein}, economic \citep{frolich2004programme}, and environmental studies \citep{zhu2024review}---subjects receive multiple treatments at the same time. For example, with two treatments $a_1$ and $a_2$, the causal relationship may take the form:
$$
E[Y^{(a_1, a_2)}] =\theta_0 +\theta_1  a_1 + \theta_2 a_2 + \theta_3 a_1 a_2.
$$
In this setting, the methods for ``multi-valued treatment'' show substantial limitations:

\begin{itemize}
    \item 	It is not straightforward to estimate $\theta_3$, the interaction effect between $a_1$ and $a_2$ on the outcome.
    \item It is particularly challenging when $a_1$ or $a_2$ is continuous. Research on continuous treatments is still being developed \citep{zhu2015boosting, kennedy2017non, kallus2018policy, brown2021propensity}, where a key difficulty is that $P(A=a) = 0$ for a specific $a$ due to its continuity. Moreover, propensity score methods may lead to worse covariate balance for continuous treatments \citep{brown2021propensity}.
\end{itemize}

In this article, we address these limitations by presenting a DML partial linear model (Section \ref{sec: DML PLR}) that accommodates (1) multiple treatments assigned simultaneously, (2) treatment interactions, and (3) treatments of varying types (binary, categorical, and continuous). To our knowledge, this is the first work to leverage machine learning methods for the robust estimation of causal effects involving both multiple treatments and their interactions.

Several previous studies examine multiple concurrent treatments, but interactions among treatments are usually not explicitly considered. \citet{siddique2019causal} studied effect estimation for multiple concurrent treatments but still treated the combination of treatments as a multi-valued categorical variable. \citet{wang2019blessings, wang2021proxy} proposed the de-confounder, focusing on causal identification for multiple treatments with unobserved confounders.  From a computer science perspective, \citet{wang2024causal} reviewed complex treatment settings, including neural network-based methods for multiple treatments \citep{qian2021estimating, zou2020counterfactual, tanimoto2021regret}.

\subsection{Structure of this paper}

Section 2 describes two settings: (1) mixed-type multiple treatments and (2) multi-valued regimens, along with notation and assumptions. Section 3 presents the DML partial linear model for estimating the effects of multiple treatments and their interactions. Section 4 presents DML interactive model for estimating the pairwise ATE of multiple regimens. Section 5 describes the Neyman orthogonality score function of DML for multiple treatments and multi-valued regimens and derives asymptotic normality.  Section 6 presents simulation studies to demonstrate the performance in different settings. Section 7 applies the DML partial linear model to study the effect of three treatments (and their interactions) on HIV-associated kidney disease. Finally, a discussion section concludes this article.

\section{Settings, notation, and assumptions}
\label{sec: notation and setting}

\subsection{Settings and notation}
\begin{figure}[H]
  \centering
  \includegraphics[width=0.75\textwidth]{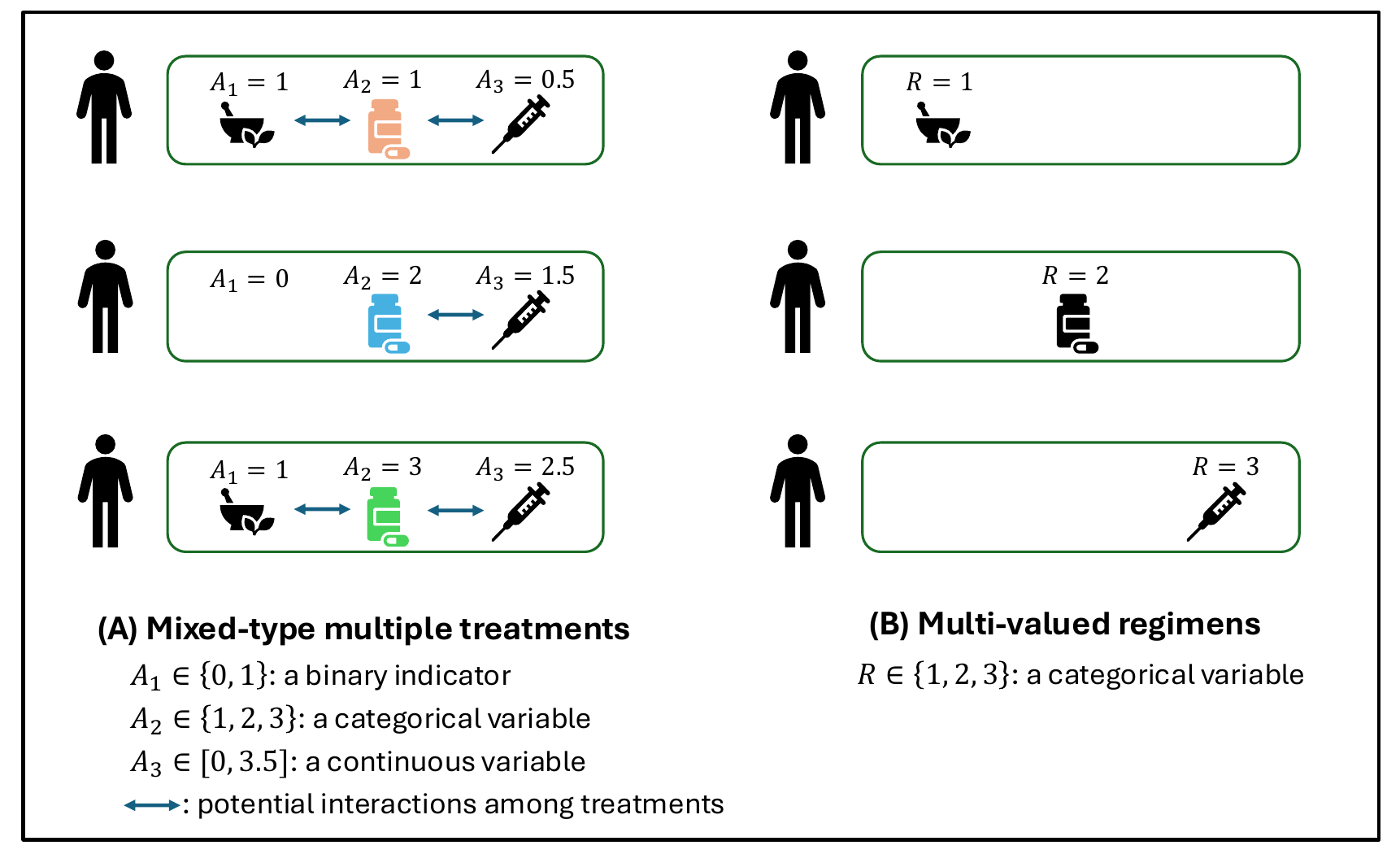}
  \caption{Illustrative settings of (A) multiple treatments and (B) multi-valued regimens. In (A), subjects simultaneously receive multiple treatments of different types, allowing for potential treatment interactions. In (B), subjects receive only one category of multi-valued regimens.}
    \label{fig: settings}
\end{figure}

We outline two settings in this article, as illustrated in Figure \ref{fig: settings}. In the first setting of multiple treatments, subject $i$ receives $D$ treatments simultaneously, denoted as a vector of treatments 
\begin{align*}
    \bm{A}_{i}=[A_{i1},A_{i2},\ldots{},A_{iD}].
\end{align*}
Here, $A_{id}$ may be a binary, categorical, or continuous treatment for $d \in \{ 1, ... ,D\}$. In addition, we explicitly model the vector of $T$ treatment interactions 
\begin{align*}
    \bm{A}^\circ_i = [ A^\circ_{i1}, A^\circ_{i2}, \dots, A^\circ_{iT}], 
\end{align*}
where $A^\circ_{it}$ represents the $t$-th interaction term, formed as the product of two or more individual treatments, e.g., $A^\circ_{i1} = A_1 A_2$, $A^\circ_{i2} = A_2 A_3$, $A^\circ_{i3} = A_1 A_2 A_3$, etc. The maximum number of interactions is $2^{D} - D - 1$, but since not all treatments necessarily interact with each other, the actual number of interactions satisfies: $0 \leq T \leq 2^{D} - D - 1$.

In the second setting of multi-valued regimens, each subject receives one regimen from $D$ regimen categories. Let $R_i = d$ be a categorical indicator of observed regimen assignment for $d \in \{1, 2, .., D \}$; that is, $R_i=d$ if subject $i$ received the $d$-th regimen. This setting is often referred to as multi-valued treatments in the causal inference literature. However, we refer to it as ``multi-valued regimen'', since it is common in practice that subjects receive not only a single level of treatment but rather a regimen, which consists of a systemic plan or a combination of treatments.

Let $Y_i$ denote the observed continuous outcome for subject $i$. $\bm{X}_i$ denotes a vector of $p$-dimensional pretreatment confounding covariates. We identify and estimate the causal effects based on the potential outcome framework \citep{rosenbaum1983central}. For multiple concurrent treatments, each subject $i$ has a set of potential outcomes, denoted as $Y_i^{(a_1,a_2,\ldots{},a_D)}$,  which represents the outcome that would have been observed had subject $i$  received the treatment combination $(A_{i1} = a_1, A_{i2} = a_2,\ldots{}, A_{iD} = a_D)$. Similarly, for multi-valued regimens, subject $i$ has $D$ potential outcomes, where $Y_i^{(r=d)}$ denotes the potential outcome under the $d$-th regimen for $d \in \{1,...,D\}$.

\subsection{Assumptions}
Under the potential outcome framework, assumptions are required to identify causal effects. For multiple concurrent treatments, we introduce the following assumptions:
\begin{itemize}
    \item Strong unconfoundedness: $Y_i^{(a_1, a_2, ..., a_D)} \ind (A_{i1}, A_{i2}, ..., A_{iD}) \rvert \bm{X}_i$, for $(a_1, a_2, ..., a_D) \in \bm{\mathcal{A}}$, where $\bm{\mathcal{A}}$ denotes the space for all possible treatment combinations.
    \item 	Positivity: $0<P((A_{i1}, A_{i2}, ..., A_{iD}) \in \bm{\mathcal{A}} \rvert \bm{X}_i )$.
\end{itemize}

Strong unconfoundedness asserts that, given confounders $\bm{X}_i$,  potential outcomes under all possible treatment combinations are independent of the assignment of multiple treatments. Positivity asserts that, given $\bm{X}_i$, there is a non-zero probability of joint distribution of the treatment vector. It enables identification of causal effects for any combination of treatments and higher-order interactions (e.g., the interaction effect of three treatments simultaneously). \citet{wang2019blessings} also considered similar assumptions.

For multi-valued regimens, since subjects only receive a single category of regimens, the assumptions are different: 
\begin{itemize}
    \item Weak unconfoundedness: $Y_i^{(r=d)} \ind 
\mathbbm{1}_{\{R_i=d\}} \rvert \bm{X}_i$, for $d \in \{1,\ldots{},D \}$.
    \item Positivity: $0<P(R_i=d \rvert \bm{X}_i )<1$ for $d \in \{1,\ldots{},D \}$.
\end{itemize}

Unlike strong unconfoundedness, weak unconfoundedness only requires the independence to be `local'---that is, within each specific regimen category denoted by $\mathbbm{1}_{\{R_i=d\}}$ \citep{imbens2000role}. Similarly, the positivity assumption for multi-valued regimens only requires the marginal probability of observing each category of the regimen is non-zero. 

\subsection{Identification and estimand}
Theorem \ref{lemma: identification} below shows identification for settings of multiple treatments:
\begin{theorem} 
\label{lemma: identification}
Under the assumptions of strong unconfoundedness and positivity, the potential outcome under multiple treatments can be identified as follows:
\begin{align*}
E[Y_i^{(a_{i1},a_{i2},\dots,a_{iD})}] = E \left[ E[Y_i \mid A_{i1} = a_{i1}, A_{i2} = a_{i2}, \dots, A_{iD} = a_{iD}, \bm{X}_i] \right].
\end{align*}
\end{theorem}
Appendix Section 1.2 proves Theorem \ref{lemma: identification}, which establishes that the potential outcome under all possible treatments can be identified from observational data. This result enables the identification of the effects of treatment interactions under the partial linear model in Section \ref{sec: DML PLR}. For example, consider an outcome model with binary treatments $a_1$ and $a_2$:
\begin{align*}
Y_i =\theta_1  A_{i1} + \theta_2 A_{i2} + \theta_3 A_{i1} A_{i2} + g(\bm{X}_i) + \varepsilon_i.
\end{align*}
where $\varepsilon_i \sim N(0,1)$.  Following Theorem \ref{lemma: identification}, the estimand for the interaction effect, $\theta_3$, can be expressed as a combination of average potential outcomes:
$
    \theta_3 = E[Y^{(a_1 = 1, a_2 = 1)}] - E[Y^{(a_1 = 0 , a_2 = 1)}]  - E[Y^{(a_1 = 1, a_2 = 0)}] + E[Y^{(a_1 = 0, a_2 = 0)}].
$
The estimands for main effects $\theta_1$ and $\theta_2$ can also be expressed as a combination of average potential outcomes. These estimands, which correspond to the coefficients in the outcome model, should be interpreted jointly, since the interaction term modifies the marginal effects of the individual treatments.

For multi-valued regimens,  relevant identification results have been previously studied under the assumptions of weak unconfoundedness and positivity, e.g., Theorem 1 in \citet{imbens2000role} and Lemma 3 in \citet{yang2016propensity}.   Though the generalized propensity score is estimated using the full sample of $\bm{X}_i$, the identification can be reached for a specific category of regimen. For example, the potential outcome under b-th regimen is identified as
$$
E[Y^{(r=b)}_i] = E[E[ Y_i | \mathbbm{1}_{\{R_i = b\}} , \bm{X}_i]].
$$
In this setting, researchers are usually interested in pairwise contrasts between regimens, for example, the ATE between the potential outcomes of the $b$-th regimen and $c$-th regimen, with $b, c$ $\in \{1, ..., D\}$. The estimand for such contrast, denoted as $ATE_{bc}$, can be expressed as:
\begin{align}
\label{eq: DML IRM estimand}
    ATE_{bc}=E[Y_i^{(r=b)}-Y_i^{(r=c)}].
\end{align}

We omit the subscript $i$ in the subsequent sections for simplicity.

\section{DML partial linear model to estimate effects of multiple concurrent treatments and their interactions}
\label{sec: DML PLR}

Consider the following data generation process. For the assignment of multiple treatments \( \bm{A} = [A_1, A_2, \dots, A_D] \), each treatment \( A_d \) has its own assignment mechanism $m_d(\bm{X})$ for $d = 1,...,D$. If $A_d$ is a continuous treatment, $m_d$ is a continuous mapping from $\bm{X}$ to $A_d$:
\begin{align}
\label{eq: DML PLR treatment}
    A_d &= m_d(\bm{X}) + \varepsilon_{A_d} ,
\end{align}
with $\varepsilon_{A_d}$ as the noise term satisfying $E[\varepsilon_{A_d} \rvert \bm{X} ] = 0$. If $A_d$ is a binary or categorical treatment, \( m_d \) is a probabilistic mapping from $\bm{X}$ to \( A_d \). For example, if $A_d$ is binary, $P(A_d = 1|\bm{X}) = m_d(\bm{X})$.

The outcome generation is given by:
\begin{align}
\label{eq: DML PLR outcome}
    Y &= \bm{A}^\top \bm{\theta} + {\bm{A}^\circ}^\top \bm{\theta}^\circ + g(\bm{X}) + \varepsilon_Y,
\end{align}
where $\bm{A}^\circ$ is a vector of $T$ interaction terms, as introduced in Section \ref{sec: notation and setting},
\begin{align*}
    \bm{A}^\circ &= [ A^\circ_1, A^\circ_2, \dots, A^\circ_T];
\end{align*}
$g(\bm{X})$ is a function mapping the confounding covariates $\bm{X}$ to $\mathbb{R}$; $\varepsilon_{Y}$ is a noise term with $E[\varepsilon_{Y} \rvert \bm{A}, \bm{A}^\circ, \bm{X} ] = 0$. $\bm{\theta}$ and $\bm{\theta}^\circ$ are the coefficients for $\bm{A}$ and $\bm{A}^\circ$, which have a causal interpretation under strong unconfoundedness, positivity, and Theorem \ref{lemma: identification}.

The treatment interactions $\bm{A}^\circ$ cannot be deterministically generated from $\bm{X}$, as the interactions are not directly assigned. However, interactions can still be confounded by  $\bm{X}$ because the individual treatment assignments are dependent on $\bm{X}$. Since interactions may affect the outcome,  they need to be included in the outcome model.

To estimate the population parameters $[\bm{\theta}^\top, \bm{\theta}^{\circ \top}]$,  we use a ``partialing-out'' procedure based on the double residual methodology \citep{robinson1988root}. This procedure includes two steps: (1) model fitting and (2) residual regression. In the first step, we fit a treatment model for each element of $\bm{A}$ and $\bm{A}^\circ$, conditional on $\bm{X}$, to obtain two vectors of residuals $\hat{\bm{\varepsilon}}_{\bm{A}}$ and $\hat{\bm{\varepsilon}}_{\bm{A^\circ}}$; we also fit an outcome model of $Y$, conditional on $\bm{X}$, to obtain the residual $ \hat{\varepsilon}_{Y|\bm{X}} $. This step can be summarized as:
\iffalse
 \begin{align*}
    \tilde{\bm{A}} = \bm{A} - \hat{E}[\bm{A} \rvert \bm{X}]  &\text{, } \tilde{\bm{A}^\circ} = \bm{A}^\circ - \hat{E}[\bm{A}^\circ \rvert \bm{X}], \\
     \tilde{Y} & = Y - \hat{E}[Y \rvert \bm{X}].
 \end{align*}
\fi
\begin{align*}
    \hat{\bm{\varepsilon}}_{\bm{A}} = \bm{A} - \hat{E}[\bm{A} \rvert \bm{X}]  &\text{, } \hat{\bm{\varepsilon}}_{\bm{A^\circ}} = \bm{A}^\circ - \hat{E}[\bm{A}^\circ \rvert \bm{X}], \\
     \hat{\varepsilon}_{Y|\bm{X}} & = Y - \hat{E}[Y \rvert \bm{X}].
\end{align*}
$\hat{\bm{\varepsilon}}_{\bm{A^\circ}}$ is calculated directly on the entire interaction term instead of computing the residuals from individual treatments separately and then multiplying them. Additionally, $\varepsilon_{Y|X}$ differs from the error term $\varepsilon_Y$ in the outcome model \eqref{eq: DML PLR outcome}, where $\varepsilon_{Y|X}$ specifically denotes the residual after conditioning only on $\bm{X}$.
 
In the second step of the ``partialling-out'' procedure, we regress the obtained residuals:
 \begin{align}
 \label{eq: residual reg}
      \hat{\varepsilon}_{Y|\bm{X}} =  \hat{\bm{\varepsilon}}_{\bm{A}}^\top \bm{\theta}  +   \hat{\bm{\varepsilon}}_{\bm{A^\circ}}^\top \bm{\theta}^\circ.
 \end{align} 
Consequently, the estimating equations of the residual regression \eqref{eq: residual reg}
will lead to an estimate of the population parameters  $[\bm{\theta}^\top, \bm{\theta}^{\circ \top}]$:
\begin{align*}
\begin{bmatrix}
    \hat{\bm{\theta}} \\
    \hat{\bm{\theta}}^\circ
\end{bmatrix}= \left( E\left[\begin{bmatrix}  \hat{\bm{\varepsilon}}_{\bm{A}} \\ \hat{\bm{\varepsilon}}_{\bm{A^\circ}} \end{bmatrix} 
\begin{bmatrix} \hat{\bm{\varepsilon}}_{\bm{A}}^\top    & \hat{\bm{\varepsilon}}_{\bm{A^\circ}}^\top \end{bmatrix} \right] \right)^{-1} 
E\left[\begin{bmatrix}  \hat{\bm{\varepsilon}}_{\bm{A}} \\ \hat{\bm{\varepsilon}}_{\bm{A^\circ}} \end{bmatrix}  \hat{\varepsilon}_{Y|\bm{X}} \right].
\end{align*}

Because of the complex confounding relationships, we use machine learning algorithms to estimate nuisance parameters $E[A \rvert \bm{X}]$, $E[A^\circ \rvert \bm{X}]$, and $E[Y \rvert \bm{X}]$.  While machine learning methods can introduce regularization bias, the ``partialling-out'' procedure ensures robust estimation for $[\bm{\theta}^\top, \bm{\theta}^{\circ \top}]$ against biases/errors in the nuisance parameter estimation. This key advantage arises from the estimating equations resulting from \eqref{eq: residual reg}, which satisfies the Neyman orthogonality condition (see Section \ref{sec: Neyman orthogonal} for more details).

\begin{figure}[H]
  \centering
  \includegraphics[width=0.75\textwidth]{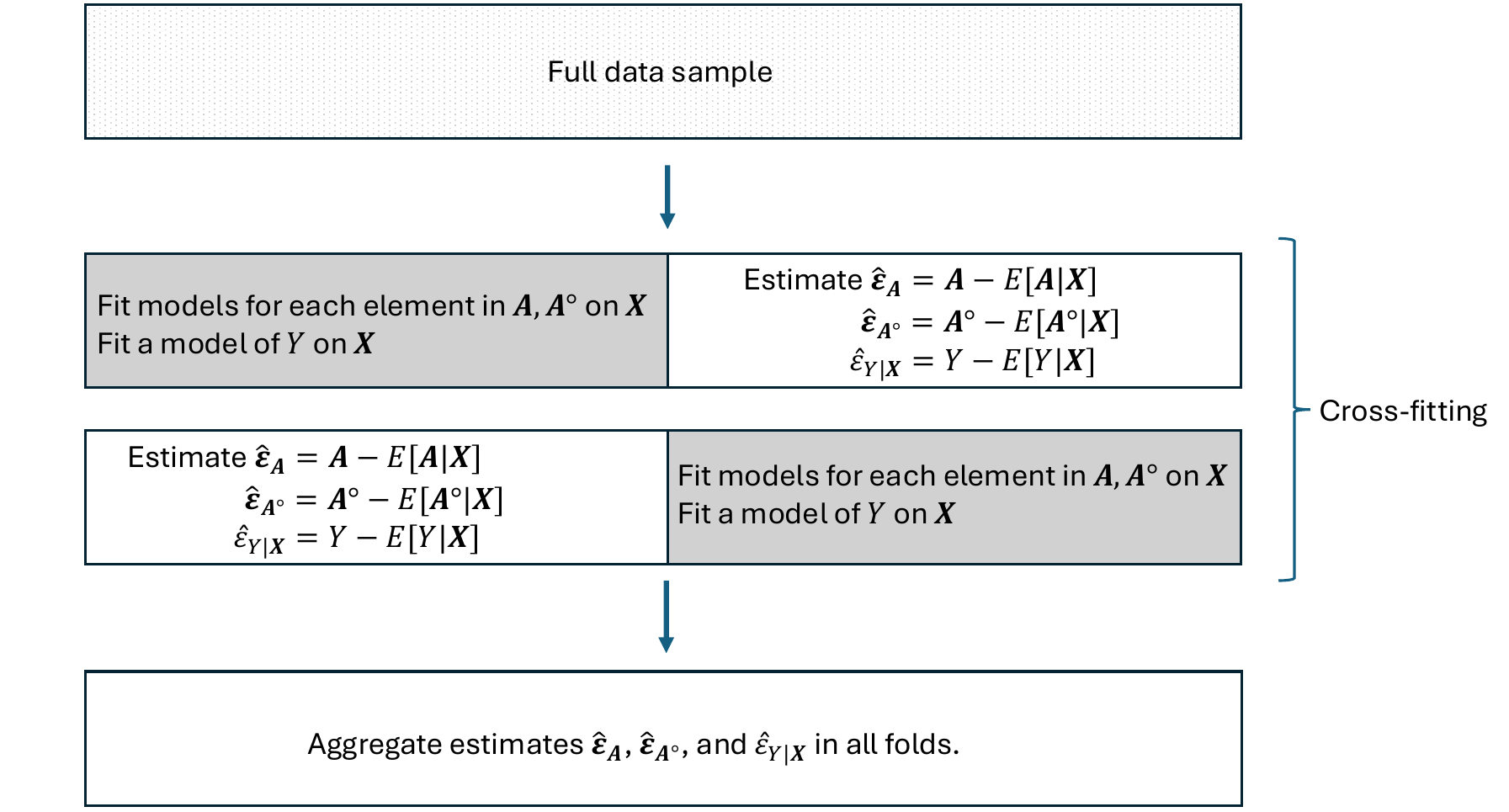}
  \caption{Illustration for 2-fold cross-fitting in settings of multiple treatments and their interactions. }
    \label{fig: cross-fitting}
\end{figure}

To overcome potential overfitting bias from machine learning algorithms, a core procedure in DML is cross-fitting (Figure \ref{fig: cross-fitting}). The data is split into $K$ folds. The machine learning models  are trained on $K-1$ folds of training data and are then used to estimate $\hat{E}[\bm{A} \rvert \bm{X}]$, $\hat{E}[\bm{A}^\circ \rvert \bm{X}]$, and $\hat{E}[Y \rvert \bm{X}]$ on the held-out fold. Therefore, the overfitting bias learned from training data does not transfer to the held-out data. Then, we calculate the residuals $\hat{\bm{\varepsilon}}_{\bm{A}}$, $\hat{\bm{\varepsilon}}_{\bm{A^\circ}}$, and $\hat{\varepsilon}_{Y|\bm{X}}$ and aggregate these residuals across all held-out folds. Finally, we plug them into the residual regression \eqref{eq: residual reg} to obtain the estimate for $[\bm{\theta}^\top, \bm{\theta}^{\circ \top}]$.

Algorithm \ref{algo: DML PLR} summarizes the procedure of DML partial linear model for estimating the effects of multiple treatments and their interactions. In Step 1.2, one can also fit a multivariate model for \( \bm{A} \) and \( \bm{A}^\circ \) on $\bm{X}$. However, this approach might be difficult when treatments $\bm{A}$ differ in types (e.g., binary, categorical, continuous). Therefore, we focus on fitting separate models for each element of \( \bm{A} \) and \( \bm{A}^\circ \).

\begin{algorithm}[H]

\caption{DML partial linear model for multiple treatments and their interactions}
\begin{algorithmic}
\State \textbf{Input:} Data $\{(\bm{X}_i, \bm{A}_i, Y_i)\}_{i=1}^N$, where $\bm{X}_i$ is a vector of covariates, $\bm{A}_i$ is the vector of multiple treatments, and $Y_i$ is the outcome. The subscript $i$ is omitted in the following steps for simplicity.

\State \textbf{Step 1: cross-fitting.}

\begin{enumerate}[label=1.\arabic*]
    \item Split the data into \( K \) folds.

    \item Fit models using \( K-1 \) folds (training data):
    \begin{itemize}
      \item Fit models for every element of treatments \( \bm{A} \) and  treatment interactions \( \bm{A}^\circ \) to obtain a series of models \( E[\bm{A} \rvert \bm{X}] \) and \( E[\bm{A}^\circ \rvert \bm{X}] \). 
        \item Fit the outcome model \( E[Y \rvert \bm{X}] \).
    \end{itemize}

    \item Use the fitted models to estimate \( \hat{E}[\bm{A} \rvert \bm{X}] \), \( \hat{E}[\bm{A}^\circ \rvert \bm{X}] \), and  \( \hat{E}[Y \rvert \bm{X}] \) on the held-out fold. Calculate the residuals:  
    \[
     \hat{\bm{\varepsilon}}_{\bm{A}}  = \bm{A} - \hat{E}[\bm{A} \rvert \bm{X}], \quad 
    \hat{\bm{\varepsilon}}_{\bm{A^\circ}}= \bm{A}^\circ - \hat{E}[\bm{A}^\circ \rvert \bm{X}], \quad 
   \hat{\varepsilon}_{Y|\bm{X}} = Y - \hat{E}[Y \rvert \bm{X}].
    \]

    \item Repeat steps 1.2 - 1.3 for all folds and aggregate the residuals \(  \hat{\bm{\varepsilon}}_{\bm{A}},  \hat{\bm{\varepsilon}}_{\bm{A^\circ}}, \hat{\varepsilon}_{Y|\bm{X}}  \) across all folds.
\end{enumerate}

\State \textbf{Step 2: estimation.} Regress on the residuals from all folds:  
\[
      \hat{\varepsilon}_{Y|\bm{X}} =  \hat{\bm{\varepsilon}}_{\bm{A}}^\top \bm{\theta}  +   \hat{\bm{\varepsilon}}_{\bm{A^\circ}}^\top \bm{\theta}^\circ.
\]
\State \textbf{Output:} The estimated coefficients, $\hat{\bm{\theta}}$ and $\hat{\bm{\theta}}^\circ$, reflect the effect of multiple treatments and their interactions on the outcome.

\end{algorithmic}
\label{algo: DML PLR}
\end{algorithm}

We now describe the residual calculation (Step 1.3 of Algorithm \ref{algo: DML PLR}) for both binary treatment and categorical treatment. For a binary treatment $A \in \{0, 1\}$, the residual is computed as 
\begin{align*}
    \hat{\varepsilon}_{A}  = A - \hat{E}[A \rvert \bm{X}] = A - \hat{p}(A \rvert \bm{X}).
\end{align*}
For a categorical treatment $A \in \{0, ... L\}$, we  estimate the conditional probabilities $\hat{p}(A=l | \bm{X})$ for each level $l \in \{0, ..., L\}$ via a multi-class classification model. We then use dummy variable encoding on $A$ to create binary indicators for each level: $A_{\text{level}: 0}$ (reference level), $A_{\text{level}: 1}$, ..., $A_{\text{level}: L}$. That is, for each level $l$, we define $A_{\text{level}: l} = 0$ if $A \neq l$  and $A_{\text{level}: l} = 1$ if $A = l$.  Residuals for each level are then computed  as 
\begin{align*}
    \tilde{A}_{\text{level}: l} = A_{\text{level}: l} - \hat{E}[A_{\text{level}: l}| \bm{X}] = A_{\text{level}: l} - \hat{p} (A = l| \bm{X}).
\end{align*}
Here, $\tilde{A}_{\text{level}: l}$ denotes the residual of $A_{\text{level}: l}$. Finally, besides the reference level $\tilde{A}_{\text{level}: 0}$, all other residuals $\tilde{A}_{\text{level}: l}$ for $l=1,...,L$ are included in the residual regression (Step 2 of Algorithm \ref{algo: DML PLR}). Therefore, the resulting coefficients  can  be interpreted as the effects of each treatment level relative to the reference level.

\section{DML interactive model to estimate ATE of multi-valued regimens}\label{sec: DML IRM}
Consider the following data generation process:

\begin{align}
P(R=d \rvert \bm{X}) & = m_d (\bm{X}) , \  d \in \{1,\ldots{},D \}, \label{eq: DML IRM regimen} \\
Y & =g(R,\bm{X}) + \varepsilon_Y,  \label{eq: DML IRM outcome}
\end{align}
where the equation \eqref{eq: DML IRM regimen} is the model of the assignment of $D$ regimens and the equation \eqref{eq: DML IRM outcome} is the outcome model. $m_d (\bm{X})$ is a function mapping the support of $X$ to the probability of the $d$-th regimen, for $d = 1,...,D$, which is also referred to as generalized propensity score.  $g(R,\bm{X})$ is a function mapping both $R$ and $\bm{X}$ to $\mathbb{R}$. $\varepsilon_Y$ is the error term with $E[\varepsilon_Y \rvert R,\bm{X}]=0$.  This ``interactive model'' allows the effects of regimens $R$ to be heterogeneous by interacting with $\bm{X}$ in the outcome model.

As shown in equation \eqref{eq: DML IRM estimand} in Section 2.3, the estimand of interest typically is the ATE between the potential outcomes of the $b$-th and $c$-th regimen, for $b,c \in \{1,..D\}$.  Based on weak unconfoundedness, positivity, and the above data generation process, there are two popular estimators for $ATE_{bc}$. First, the regression adjusted estimator using the outcome model, 
\begin{align*}
    \widehat{ATE}_{bc}= \mathbb{E}_n [\hat{g}(r=b,\bm{X})-\hat{g}(r=c,\bm{X})],
\end{align*}
and second, the IPW estimator using the generalized propensity score
\begin{align*}
\widehat{ATE}_{bc}=\mathbb{E}_n\left[\frac{\mathbbm{1}_{\{R=b\}}}{\hat{m}_b(\bm{X})} Y - \frac{\mathbbm{1}_{\{R=c\}}}{\hat{m}_c(\bm{X})} Y \right].
\end{align*}
Here, $\mathbb{E}_n$ denotes the empirical average, i.e., $\mathbb{E}_n\left[f(X_i)\right] = \frac{1}{n} \sum_{i=1}^n f(X_i)$. However, if the outcome model $g$ or the generalized propensity score model $m$ is misspecified, these estimators can be biased and inefficient. 

To mitigate the limitations of the previous estimators, a doubly robust approach combines both outcome and propensity score models \citep{robins1995analysis, bang2005doubly}, which ensures the estimate is consistent if either model is correctly specified. This estimator for multi-valued regimens is given by:

\begin{align}
    \label{eq: DML IRM estimator}
    & \widehat{ATE}_{bc} =  \\  \nonumber
    &  \mathbb{E}_n\left[ \big[\hat{g}(r = b, \bm{X}) - \hat{g}(r = c, \bm{X})\big]  + \frac{\mathbbm{1}_{\{R=b\}} \big(Y - \hat{g}(r = b, \bm{X})\big)}{\hat{m}_b(\bm{X})}  - \frac{\mathbbm{1}_{\{R=c\}} \big(Y - \hat{g}(r = c, \bm{X})\big)}{\hat{m}_c(\bm{X})} \right].
\end{align}

\iffalse

As we mentioned in the introduction, the covariate-treatment relationship and covariate-outcome relationship and can be complicated, for example, the nonlinear relationships. Additionally, the presence of high-dimensional covariates $X$ further complicates estimation. So it comes naturally to use machine learning methods, who is advantage in nonlinear relationship and high-dimensional covariates, to estimate $m$ and $g$. However, machine learning methods are subject to overfitting or regularization bias, since the goal of machine learning is almost prediction not estimation. 
\fi

The estimating equation from \eqref{eq: DML IRM estimator} satisfies the Neyman orthogonal condition. Similar to Section \ref{sec: DML PLR}, we use machine learning methods to estimate $m$ and $g$ with cross-fitting. The Algorithm \ref{algo: DML IRM} summarizes the whole procedure.

\begin{algorithm}[H]

\caption{DML interactive model for multi-valued regimens}
\begin{algorithmic}
\State \textbf{Input:} Data $\{(\bm{X}_i, R_i, Y_i)\}_{i=1}^N$, where $\bm{X}_i$ is a vector of  covariates, $R_i$ is the observed multi-valued regimen, and $Y_i$ is the outcome. The subscript $i$ is omitted in the following steps for simplicity.

\State \textbf{Step 1: cross-fitting.}

\begin{enumerate}[label=1.\arabic*]
    \item Split the data into \( K \) folds.

    \item Fit models using \( K-1 \) folds (training data):
    \begin{itemize}
        \item Fit a multivariate model for generalized propensity score: $P(R = d \mid \bm{X}) = m_d(\bm{X}), \ d \in \{1,\ldots{},D \}$.
        \item Fit the outcome models $Y = g(R=b, \bm{X})$ and $Y = g(R=c, \bm{X})$ conditioning on $R = b$ and $R=c$, respectively.
    \end{itemize}

    \item Estimate using the $\bm{X}$ in the held-out fold:
    \begin{itemize}
        \item Estimate $\hat{m}_b(\bm{X})$ (the propensity of receiving the $b$-th regimen) and $\hat{m}_c(\bm{X})$ (the propensity of receiving the $c$-th regimen).
        \item  Estimate $\hat{g}(r = b, \bm{X})$ and $\hat{g}(r = c, \bm{X})$ for the $b$-th and $c$-th regimens, respectively.
    \end{itemize}
    
    \item Repeat the steps 1.2 - 1.3 for all folds, and aggregate the estimates $\hat{m}_b(\bm{X})$,  $\hat{m}_c(\bm{X})$, $\hat{g}(r = b, \bm{X})$, and $\hat{g}(r = c, \bm{X})$ across all folds.
\end{enumerate}

\State \textbf{Step 2: estimation.}  Use the doubly robust estimator in equation \eqref{eq: DML IRM estimator} to estimate the ATE:
$$    
    \widehat{ATE}_{bc} = \mathbb{E}_n \left[ \big[\hat{g}(r = b, \bm{X}) - \hat{g}(r = c, \bm{X})\big] 
    + \frac{\mathbbm{1}_{\{R=b\}} \big(Y - \hat{g}(r = b, \bm{X})\big)}{\hat{m}_b(\bm{X})} 
    - \frac{\mathbbm{1}_{\{R=c\}} \big(Y - \hat{g}(r = c, \bm{X})\big)}{\hat{m}_c(\bm{X})} \right].
$$ 
\State \textbf{Output:} $\widehat{ATE}_{bc}$, the average treatment effect between the $b$-th regimen and $c$-th regimen.

\end{algorithmic}
\label{algo: DML IRM}
\end{algorithm}

\section{Neyman orthogonality score function and variance estimation}
\label{sec: Neyman orthogonal}

This section discusses the score function for multiple treatments and multi-valued regimens. We also describe the asymptotic normality for the estimators in both settings and thereby derive the variance estimator.

\subsection{Neyman orthogonality score}
For our proposed model, we consider the method of moments estimator:
\begin{align*}
    E[\psi(\bm{W}; \bm{\Theta}_0, \bm{\eta}_0)] = 0,
\end{align*}
where $\psi$ is a vector-valued score function that characterizes the moment conditions; $\bm{W}$ denotes the sample space: $(\bm{X}, Y, \bm{A})$ for multiple treatments and $(\bm{X}, Y, R)$ for multiple regimens; $\bm{\Theta}_0$ denotes the true value of parameters of interest, e.g., for multiple treatments and interactions, $\bm{\Theta}_0^{\top} = [\bm{\theta}^\top, \bm{\theta}^{\circ \top}]$;  $\bm{\eta}_0$ denotes the true values of the nuisance parameters; 

The \textit{Neyman orthogonality} condition requires the score function $\psi$ satisfies:
\begin{align}
    \partial_t E[\psi(\bm{W}; \bm{\Theta}_0, \bm{\eta}_0 + t\bm{\delta})]\big|_{t=0} = 0,
\label{eq: Neyman condition}
\end{align}
where $\bm{\delta} = \bm{\eta} - \bm{\eta}_0$ represents deviations of the nuisance parameters from their true values. This condition ensures that the moment condition evaluated at $\bm{\Theta}_0$ are not sensitive to the small bias in estimating $\bm{\eta}_0$, thereby preserving the unbiased estimation for $\bm{\Theta}_0$.

In this study, the score function for both DML partial linear model and interactive model is linear in $\bm{\Theta}$, and thereby can be expressed as 
\begin{align}
    \label{eq: Neyman linear decom}
    \psi(\bm{W}; \bm{\Theta}, \bm{\eta}) = \psi_a (\bm{W}; \bm{\eta}) \bm{\Theta} +  \psi_b (\bm{W}; \bm{\eta}),
\end{align}
where $\psi_a (\bm{W}; \bm{\eta})$ is the  component linear in $\theta$ and $\psi_b (\bm{W}; \bm{\eta})$ is independent of $\theta$. Based on this linear representation, we can also estimate $\bm{\Theta}$ by solving the moment condition:
\begin{align}
\label{eq: estimator from linear score}
     \bm{\Theta} & = -E [ \psi_a (\bm{W}; \bm{\eta}) ]^{-1} E[\psi_b (\bm{W}; \bm{\eta})].
\end{align}
The linearity representation in Equation \ref{eq: Neyman linear decom} can facilitate the estimation of the variance, which will be presented in the following subsections.

\subsection{Neyman orthogonality and variance estimation for multiple treatments and their interactions}

In this setting, the score function is formulated based on the ``partialling-out'' approach described in Section \ref{sec: DML PLR}. Lemma \ref{lemma: PLR score function} shows  that such score function satisfies the Neyman orthogonality condition in \eqref{eq: Neyman condition}.

\begin{lemma}{}
The score function of the DML partial linear model for multiple treatments and interactions,
\begin{align}
\label{eq: PLR score function}
    \psi(\bm{W}; \bm{\Theta}, \bm{\eta}) &= \begin{bmatrix} \bm{A} -\bm{m}(\bm{X})\\ \bm{A}^\circ - \bm{m}^\circ(\bm{X}) \end{bmatrix} \left( Y - l(\bm{X}) -  \begin{bmatrix} \bm{A} -\bm{m}(\bm{X})\\ \bm{A}^\circ - \bm{m}^\circ(\bm{X}) \end{bmatrix}^\top \begin{bmatrix}
        \bm{\theta} \\
        \bm{\theta}^{\circ}
    \end{bmatrix} \right),
\end{align}
satisfies the Neyman orthogonality condition.
\label{lemma: PLR score function}
\end{lemma}

In this lemma, $\bm{\Theta}_0^{\top} = [\bm{\theta}^\top, \bm{\theta}^{\circ \top}]$ and the nuisance parameters $\bm{\eta} = (\bm{m}, \bm{m}^\circ, l)$, i.e., $\bm{m}(\bm{X}) = E[\bm{A} | \bm{X}]$, $\bm{m}^\circ(\bm{X}) = E[\bm{A}^\circ | \bm{X}]$, and $l(\bm{X}) = E[Y | \bm{X}]$. The Appendix Section 1.2 proves the lemma \ref{lemma: PLR score function}.

Based on this score function,  we can also estimate the parameter of interest following equation \eqref{eq: Neyman linear decom} and \eqref{eq: estimator from linear score}:
\begin{align*}
     \begin{bmatrix}
        \bm{\theta} \\
        \bm{\theta}^{\circ}
    \end{bmatrix} & = -E(\psi_a)^{-1} E(\psi_b)  \\
     & = E \left[ \begin{bmatrix} \bm{A} -\bm{m}(\bm{X})\\ \bm{A}^\circ - \bm{m}^\circ(\bm{X}) \end{bmatrix} \begin{bmatrix} \bm{A} -\bm{m}(\bm{X})\\ \bm{A}^\circ - \bm{m}^\circ(\bm{X}) \end{bmatrix}^\top \right]^{-1} E\left[ \begin{bmatrix} \bm{A} -\bm{m}(\bm{X})\\ \bm{A}^\circ - \bm{m}^\circ(\bm{X}) \end{bmatrix} (Y - l(\bm{X}))  \right] \\
    & =
     E\left[\begin{bmatrix}  \bm{\varepsilon}_{\bm{A}} \\ \bm{\varepsilon}_{\bm{A^\circ}} \end{bmatrix} 
\begin{bmatrix} \bm{\varepsilon}_{\bm{A}}^\top    & \bm{\varepsilon}_{\bm{A^\circ}}^\top \end{bmatrix} \right] ^{-1} 
E\left[\begin{bmatrix}  \bm{\varepsilon}_{\bm{A}} \\ \bm{\varepsilon}_{\bm{A^\circ}} \end{bmatrix}  \bm{\varepsilon}_{Y| \bm{X}} \right].
\end{align*}
The third equality in the equation above uses that $ \bm{A} -\bm{m}(\bm{X}) = \bm{\varepsilon}_{\bm{A}}$ and $ Y - l(\bm{X}) =\bm{\varepsilon}_{Y|\bm{X}} $. Note that such estimating equation leads to an estimator of the same format as the one derived using the residual regression in Section \ref{sec: DML PLR}.

Proposition \ref{thm: DML PLR} introduces the asymptotic normality of this estimator for $[\bm{\theta}^\top, \bm{\theta}^{\circ \top}]$.

\begin{proposition}[Asymptotic normality of DML for multiple treatments and interactions]\label{thm: DML PLR} 
Under regularity conditions in the Appendix, the estimator $[\hat{\bm{\theta}}^\top, \hat{\bm{\theta}}^{\circ \top} ]$  satisfies:
\[
\sqrt{n} \left( \begin{bmatrix}
    \hat{\bm{\theta}} \\
    \hat{\bm{\theta}}^\circ
\end{bmatrix}  - \begin{bmatrix}
    \bm{\theta}_0 \\
    \bm{\theta}^\circ_0
\end{bmatrix} \right) \overset{d}{\to} N\left( \bm{0}, \, \bm{\Sigma} \right),
\]
with asymptotic variance-covariance matrix
\begin{align}
\label{eq: PLR variance}
    \bm{\Sigma} =  \bm{J}_0^{-1} E\left[\psi(\bm{W}; \bm{\Theta}_0, \bm{\eta}_0) \psi(\bm{W}; \bm{\Theta}_0, \bm{\eta}_0)^\top \right] (\bm{J}_0^{-1})^\top,
\end{align}
where
\begin{align*}
       \bm{J}_0 = E[\psi_a(\bm{W}, \bm{\eta}_0)] = -E \left[\begin{bmatrix}
        \bm{\varepsilon_A} \\
        \bm{\varepsilon_{A^\circ}}
    \end{bmatrix}  \begin{bmatrix}
        \bm{\varepsilon_A}^\top \; \bm{\varepsilon_{A^\circ}}^\top
    \end{bmatrix}   \right].
\end{align*}
\end{proposition}

As a consequence of Proposition \ref{thm: DML PLR}, after computing the estimated score function $\hat{\psi}(\bm{W}; \hat{\bm{\Theta}}, \hat{\bm{\eta}})$ and  $\hat{\bm{J}}$, we can estimate the variance using formula \eqref{eq: PLR variance}:

\begin{align*}
    \hat{\bm{\Sigma}} & =  \hat{\bm{J}}^{-1} E\left[\hat{\psi}(\bm{W}; \hat{\bm{\Theta}}, \hat{\bm{\eta}}) \hat{\psi}(\bm{W}; \hat{\bm{\Theta}}, \hat{\bm{\eta}})^\top \right] (\hat{\bm{J}}^{-1})^\top \\
    &=  E\left[\begin{bmatrix}  \hat{\bm{\varepsilon}}_{\bm{A}} \\ \hat{\bm{\varepsilon}}_{\bm{A^\circ}} \end{bmatrix} 
\begin{bmatrix} \hat{\bm{\varepsilon}}_{\bm{A}}^\top    & \hat{\bm{\varepsilon}}_{\bm{A^\circ}}^\top \end{bmatrix} \right]^{-1}
    E \left[ \hat{\varepsilon}_Y  \begin{bmatrix}  \hat{\bm{\varepsilon}}_{\bm{A}} \\ \hat{\bm{\varepsilon}}_{\bm{A^\circ}} \end{bmatrix} 
\begin{bmatrix} \hat{\bm{\varepsilon}}_{\bm{A}}^\top    & \hat{\bm{\varepsilon}}_{\bm{A^\circ}}^\top \end{bmatrix} \hat{\varepsilon}_Y \right] \left(E\left[\begin{bmatrix}  \hat{\bm{\varepsilon}}_{\bm{A}} \\ \hat{\bm{\varepsilon}}_{\bm{A^\circ}} \end{bmatrix} 
\begin{bmatrix} \hat{\bm{\varepsilon}}_{\bm{A}}^\top    & \hat{\bm{\varepsilon}}_{\bm{A^\circ}}^\top \end{bmatrix} \right]^{-1} \right)^\top.
\end{align*}

\subsection{Neyman orthogonality and variance estimation for multi-valued regimens}

In this setting, we used the DML interactive model to estimate the ATE between two regimen categories (Section \ref{sec: DML IRM}), and we introduce Lemma \ref{lemma: IRM score function}

\begin{lemma}
The score function of the DML interactive model to estimate the ATE,
\begin{align}
    &\psi(\bm{W}; \theta, \bm{\eta})  = \big[g(r = b, \bm{X}) - g(r = c, \bm{X})\big]  + \\
     & \qquad \qquad \frac{\mathbbm{1}_{R=b} \big(Y - g(r = b, \bm{X})\big)}{m_b(\bm{X})}  - \frac{\mathbbm{1}_{R=c} \big(Y - g(r = c, \bm{X})\big)}{m_c(\bm{X})} - \theta, \nonumber
\end{align}
satisfies the Neyman orthogonality condition.
\label{lemma: IRM score function}
\end{lemma}

Appendix Section 1.4 proves the Lemma \ref{lemma: IRM score function}. In this lemma, $\theta$ denotes the parameter of interest: ATE between the $b$-th regimen and the $c$-th regimen, $b, c \in {1, ..., D}$. Note that the Neyman orthogonal score function in this settings presents double robustness \citep{bang2005doubly}.  Based on this Neyman orthogonal score, Proposition \ref{thm: DML IRM} below introduces the asymptotic normality.
 
\begin{proposition}[Asymptotic Normality for DML with multi-valued treatments]\label{thm: DML IRM} 
Under regularity conditions in the Appendix, the estimator $\hat{\theta}$  satisfies:
\[
\sqrt{n} \left( \hat{\theta} - \theta_0 \right) \overset{d}{\to} N\left( 0, \, \sigma^2 \right),
\]
with asymptotic variance
\begin{align}
\label{eq: IRM variance}
    \sigma^2 =  E\left[\psi(\bm{W}; \theta_0, \bm{\eta}_0)^2 \right].
\end{align}
\end{proposition}

As a consequence of Proposition \ref{thm: DML IRM}, after computing the estimated score function $\hat{\psi}(\bm{W}; \hat{\theta}, \hat{\bm{\eta}})$, we can estimate the variance using formula \eqref{eq: IRM variance}: 
\begin{align*}
    \hat{\sigma}^2 & =  E\left[\hat{\psi}(\bm{W}; \hat{\theta}, \hat{\bm{\eta}})^2 \right] 
\end{align*}

\section{Simulation studies}
\subsection{Simulation for DML partial linear model}

This subsection evaluates the performance of the DML partial linear model on a simulated dataset. We generate covariates $\bm{X}$ of 10 dimensions. Specifically, covariates $X_1$ to $X_5$ are drawn independently from $N(0,1)$, and $X_6$ to $X_{10}$ are drawn independently from a Bernoulli distribution with different probabilities of success. We generate two treatments:

\begin{itemize}
    \item a binary treatment generated as $A_1 \sim \text{Bernoulli} (\pi_1)$, where  $\text{logit}(\pi_1) = m_1(\bm{X})$. 
    \item a continuous treatment $A_2$ generated by $A_2 = m_2(\bm{X}) + \varepsilon_{A_2}$, where $ \varepsilon_{A_2} \sim N(0, 1)$.
\end{itemize}
The functions $m_1(\bm{X})$ and $m_2(\bm{X})$  incorporate a mixture of linear, nonlinear, and interaction effects of the covariates:
\begin{align*}
    m_1(\bm{X}) & = 1.3 X_1 X_2 + 0.7 X_2^2 - 0.4 X_3 + e^{X_4} + 1.5 X_7 X_9 - 1.5 X_{10}. \\
    m_2(\mathbf{X}) & = \frac{1}{1 + e^{X_1}} - \frac{1}{1 + e^{X_2}} + 0.5 X_3 + 0.25 \left( \mathbbm{1}(X_5 > 0) - \mathbbm{1}(X_6 > 0) \right) + 0.1 \left( X_7 + X_9 X_{10} \right).
\end{align*}

The outcome $Y$ is generated by a model of these two treatments and their interaction, as well as other covariates:
\begin{align*}
    Y= \theta_{1}  A_1 + \theta_{2}   A_2+ \theta_{3}  A_1  A_2 + g(\bm{X}) + \varepsilon,
\end{align*}
where 
\begin{align*}
    g(\bm{X}) =  \bm{X}^L \bm{\beta}^L_d + \bm{X}^{NL} \bm{\beta}^{NL}_d & =  -2 \mathbbm{1}(X_1 < 0) +  2\mathbbm{1}(X_1 \geq 0)  -  \mathbbm{1}(X_2 < 1) +  \mathbbm{1}(X_2 \geq 1) \\
& \qquad + 2X_3 + 2X_5 + X_6 + X_7 - 2X_9 - 0.5X_{10} \\
& \qquad + 2X_3 X_4 + 2X_5 X_{10} + 2X_5^2 + 2X_9^2,
\end{align*}
and $\varepsilon$ is a noise term of normal distribution $N(0,1)$.  The true effects for treatments are set as
\begin{align*}
    \theta_{1} = 4 \text{, } \theta_{2} = 6 \text{, and } \theta_{3} = 4.
\end{align*}

We use four different methods to estimate the treatment models and the outcome model: an ordinary regression method, two tree-based methods (random forest and boosting trees) and a neural network method. The details of parameter choices and model specification are provided in Appendix Section 2.1. To evaluate whether these methods can capture the complex pattern of confounding covariates, we input the original covariate matrix $\bm{X}$ without transformations. 

\begin{table}[t]
\centering

\renewcommand{\arraystretch}{1.1}
\begin{tabular}{lccccc}
\toprule
  & & \makecell{DML \\Regression} & \makecell{DML \\ Random Forest} & \makecell{DML \\ Boosting Trees} & \makecell{DML \\ Neural Network}   \\
\midrule
\multicolumn{2}{l}{2-fold cross-fitting} & \multicolumn{4}{l}{} \\
\multirow{2}{*}{$\theta_1$} 
    & Bias & 0.09   & 0.04 & -0.04 & -0.07 \\
    & rMSE & 0.29   & 0.17 & 0.19 & 0.24 \\
    & CP  &  93.6\%  &    98.7\%   &  97.0\%    &  98.9\%     \\

\addlinespace[0.5em]
\multirow{2}{*}{$\theta_2$}
    & Bias & -0.36 & -0.11 & -0.09 & -0.03 \\
    & rMSE & 0.41 & 0.16 & 0.14 & 0.16 \\
    & CP  &  50.8\%   &   93.8\%   &   96.2\%   &    98.8\%    \\

\addlinespace[0.5em]
\multirow{2}{*}{$\theta_3$}
    & Bias & 0.52 & 0.12 & 0.06 & -0.01 \\
    & rMSE & 0.58 & 0.19 & 0.16 & 0.21 \\
    & CP  &   38.2\%   &   95.0\%   &   97.0\%   &   97.4\%    \\

\midrule
\multicolumn{2}{l}{5-fold cross-fitting} & \multicolumn{4}{l}{} \\

\multirow{2}{*}{$\theta_1$} 
    & Bias & 0.10 & 0.03 & -0.04 & 0.03 \\
    & rMSE & 0.29 & 0.18 & 0.19 & 0.23 \\
    & CP  &  93.4\%    &  98.7\%    &  96.6\%    &   98.6\%    \\

\addlinespace[0.5em]
\multirow{2}{*}{$\theta_2$} 
    & Bias & -0.36 & -0.11 & -0.09 & 0.03 \\
    & rMSE & 0.41 & 0.16 & 0.15 & 0.16 \\
    & CP  &   51.0\%   &    91.0\%  &  93.8\%    &    98.2\%   \\

\addlinespace[0.5em]
\multirow{2}{*}{$\theta_3$} 
    & Bias & 0.52 & 0.10 & 0.11 & -0.08 \\
    & rMSE & 0.58 & 0.17 & 0.18 & 0.23 \\
    & CP  &   38.6\%   &  95.0\%    &   93.8\%   &   92.2\%    \\

\bottomrule
\end{tabular}

\caption{DML partial linear models for multiple treatments and their interactions, with different machine learning algorithms to fit both treatment and outcome models. $\theta_1$: the effect of $A_1$.  $\theta_2$: the effect of $A_2$. $\theta_3$: the effect of the interaction of $A_1$ and $A_2$.}
\label{tab: sim DML PLR results}
\end{table}

We use Algorithm \ref{algo: DML PLR} to estimate $\theta_1$, $\theta_2$ and $\theta_3$, where we consider both 2-fold and 5-fold cross-fitting in our algorithm. We generate 500 datasets with a sample size of $n=1000$. Since the estimates can vary depending on the random sample splits in cross-fitting, we make 50 different splits on this dataset and thereby run 50 estimations. We use the median estimate and median standard error across 50 splits as the final estimates.  

Table \ref{tab: sim DML PLR results} summarizes the simulation results. Based on these results:

\begin{itemize}
    \item Regarding the bias for $\theta_1$, $\theta_2$, and $\theta_3$, tree-based models and the neural network show much lower bias than ordinary regression. This highlights that the DML using machine learning models can better capture complex relationships in this simulated data of multiple treatments.

    \item Regarding the rMSE, ordinary regression shows the highest rMSEs, consistent with its higher bias. For all methods, the rMSE for $\theta_1$ is relatively higher, suggesting that estimation of the effect of binary treatments shows more variability.

    \item Regarding the coverage probability, the tree-based models are around 0.95, while neural network model shows higher coverage probabilities. All methods have a higher coverage for $\theta_1$, suggesting the DML confidence interval for binary treatments can be conservative. Previous research  \citep{jiang2025double, yang2020double} also reported that DML confidence intervals can be conservative in certain settings.
    
    \item Results between 2-fold and 5-fold cross-fitting are comparable. However, for the interaction term $\theta_3$, boosting trees and the neural network show higher bias under 2-fold cross-fitting.
\end{itemize}

\iffalse

Table X summarizes the empirical variance, estimated analytical variance, and the coverage probability of the simulations results.

\fi

\subsection{Simulations for DML interactive model} 
Consider the following data generation process of covariates $\bm{X}$ of 10 dimensions and regimens $R$ of three categories. Covariates $X_1$ to $X_5$ follow a normal distribution of $N(0,1)$ and $X_6$ to $X_{10}$ follow a binomial distribution with different probabilities of success. $R=1,2,$ or $3$, is generated by a multinomial/softmax regression model:
\begin{align*}
    R \sim \text{Multinomial}(N, m_1(\bm{X}), m_2(\bm{X}), m_3(\bm{X})),
\end{align*}
with the probability 
\[
m_d(\bm{X}) = \frac{e^{l_d(\bm{X})}}{ \sum_{d=1}^{3} e^{l_d(\bm{X})} },
\]
where $l_1(\bm{X}) = 1$ and $l_d(\bm{X}) = \bm{X}^L \beta^L_d + \bm{X}^{NL} \beta^{NL}_d$ for $d=2,3$. See detailed model specification for \( l_d(\bm{X}) \) in the appendix. In this setting, \( R= 1 \) can be treated as a control regimen due to \( l_1(\bm{X}) = 1 \), and the relationships between regimens are
\begin{align*}
  \ln\left(\frac{P(R = 2)}{P(R = 1)}\right) &= l_2(\bm{X}) \text{ and }
        \ln\left(\frac{P(R = 3)}{P(R = 1)}\right) = l_3(\bm{X}).
\end{align*}

$Y$ is a continuous outcome generated by the outcome model $g(R,\bm{X})$ that allows for the interaction between $R$ and $\bm{X}$: 
\begin{align*}
   Y = g(R, \bm{X}) = 5 \cdot \mathbbm{1}_{R=2}+15\cdot \mathbbm{1}_{R=3} \cdot X_9 + \bm{X}^L \bm{\beta}^L_d + \bm{X}^{NL} \bm{\beta}^{NL}_d + \varepsilon,
\end{align*}
where $\mathbbm{1}_{R=3}$ interacts with $X_9$.  The causal estimand is the pairwise ATE: 
$$
ATE_{21} = 5, ATE_{31} = 10.5, \text{and } ATE_{32} = 5.5.
$$

\begin{figure}[H]
  \centerline{\includegraphics[width=5.0 in]{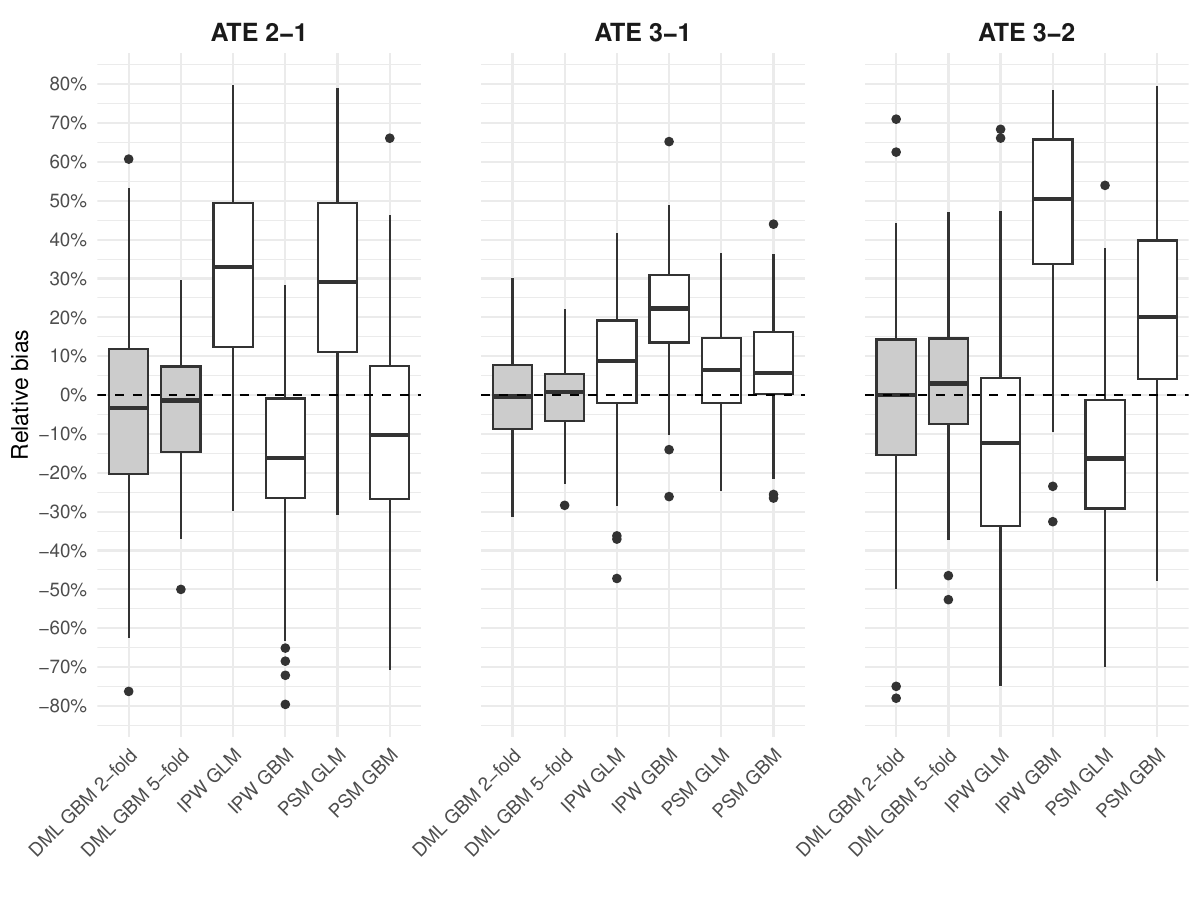}}
  \caption{Relative bias of three methods for multi-valued regimens: double machine learning (DML), inverse probability weighting (IPW), and  propensity score matching (PSM). In DML, we show results for both 2-fold cross-fitting and 5-fold cross-fitting. In IPW and PSM, the propensity scores are estimated by both generalized linear models (GLM) and gradient boosting machine (GBM). The true treatment effects: ATE\textsubscript{2-1} = 5, ATE\textsubscript{3-1} = 10.5, ATE\textsubscript{3-2} = 5.5. }
\label{fig: multi-val trt sims}
\end{figure}

We compare the performance of DML interactive model with two other previously proposed methods based on the generalized propensity score: IPW and PSM. We use the Python package \texttt{DoubleML} to implement DML interactive model \citep{bach2022doubleML}, where both the outcome and treatment models are fitted using boosting trees.. We use the R package \texttt{PSweight} \citep{zhout2022PSweight} to implement IPW and PSM, where the generalized propensity score is estimated using both the Generalized Linear Models (GLMs) and the boosting trees. 

The performance of all methods is measured by relative bias. All methods use the original covariate matrix $\bm{X}$ as the input. 100 datasets are generated with $N=1000$. For each dataset, we run 50 splits of cross-fitting in DML and use the median estimates to report the results. More details are included in Appendix.

The Figure \ref{fig: multi-val trt sims} shows the simulation results. The DML interactive model with boosting trees shows the lowest relative bias across all ATE comparisons. In addition, the 5-fold cross-fitting presents slightly less variability of relative bias than the 2-fold cross-fitting. In contrast, traditional IPW and PSM methods show larger relative biases and greater variability, even when the propensity score is estimated using boosting trees. These findings suggest that DML interactive model, which leverages double robustness by modeling both treatment and outcome, provides more robust estimates for ATE of multi-valued regimens.

\section{Application}
\label{sec: appication}

We apply the DML partial linear model to  estimate the effect of three treatments on HIV-associated kidney disease in an adult HIV cohort in Nigeria. This is a cross-sectional study conducted at Aminu Kano Teaching Hospital in northern Nigeria between 2018 and 2019.  The first visit (baseline) collected  demographic and clinical information from participants. The  final visit (5-9 weeks after baseline) collected blood samples to assess the estimated glomerular filtration rate (eGFR), a measure for kidney function.

Table \ref{tab: treatment info} summarizes the usage information of  three treatments in this study:
\begin{enumerate}
 \item All participants are on one of three types of antiretroviral therapy (ART) based on different core agents: Non-Nucleoside Reverse Transcriptase Inhibitor (NNRTI), boosted Protease Inhibitor (bPI), or Dolutegravir (DTG). This is denoted by a three-class categorical variable  $A_{\text{ART}}$. 
\item Some patients are using Tenofovir Disoproxil Fumarate (TDF) while others are not.   This is denoted by a binary variable $A_{\text{TDF}}$.
\item Some patients are using anti-hypertensive (anti-HTN) medications while others are not. This is denoted by a binary variable $A_{\text{HTN}}$.
\end{enumerate}
Because the number of participants on anti-HTN medications is small, it is difficult to estimate interactions between anti-HTN medications and other treatments. We will focus on the interaction between ART and TDF, i.e., $A^\circ_{\text{ART*TDF}}$.

\begin{table}[H]
\centering
\begin{tabular}{llrrrr}
\toprule
 & & \makecell{\textbf{ART (NNRTI)} \\ $N = 1435$}  & \makecell{\textbf{ART (bPI)} \\ $N = 431$} & \makecell{\textbf{ART (DTG)} \\$N = 589$}  \\
\midrule
\multirow{2}{*}{\textbf{Off TDF} }
  & Off anti-HTN & 732 (30\%) & 123 (5.0\%) & 15 (0.6\%)  \\
  & On anti-HTN  & 126 (5.1\%) & 11 (0.4\%) & 3 (0.1\%)  \\
\midrule
\multirow{2}{*}{\textbf{On TDF} } 
  & Off anti-HTN & 540 (22\%) & 268 (11\%) & 498 (20\%)  \\
  & On anti-HTN  & 37 (1.5\%) & 29 (1.2\%) & 73 (3.0\%)  \\
\bottomrule
\end{tabular}

    \caption{This table presents number and percentage of the treatment usage under different treatment combinations. The columns are stratified by the three-class ART treatment based on different core agent: Non-Nucleoside Reverse Transcriptase Inhibitor (NNRTI), boosted Protease Inhibitor (bPI), or Dolutegravir (DTG). The rows are stratified by the usage of Tenofovir Disoproxil Fumarate (TDF) and anti-hypertensive (anti-HTN) medications.}
    \label{tab: treatment info}
\end{table}

We use $Y_{\text{eGFR}}$ to denote the outcome eGFR, a measure for chronic kidney disease, with higher eGFR values indicating better kidney function. The confounding covariates $\bm{X}$ include: age, sex, ethnicity (Hausa-Fulani, Igbo, Yoruba, or other), risk allele (Apolipoprotein-1), diabetes mellitus status, hypertension status, congestive heart failure status, other comorbid condition status,  smoking status, body mass index, systolic and diastolic blood pressure.

\subsection{Modeling}

We use the DML partial linear model in Algorithm \ref{algo: DML PLR} to estimate the effects of  $A_{\text{ART}}$, $A_{\text{TDF}}$, $A_{\text{HTN}}$, and the interaction term $A^\circ_{\text{ART*TDF}}$ , on the outcome $Y_{\text{eGFR}}$. Since $A_{\text{ART}}$ is a three-class categorical variable and $A^\circ_{\text{ART*TDF}}$ is a six-class categorical variable, we use the residual calculation for categorical treatment described in Section \ref{sec: DML PLR}. For $A_{\text{ART}}$, we fit a multi-class classification model to estimate the conditional probability of each class of ART treatment. We then use dummy variable encoding on $A_{\text{ART}}$ to create three binary dummy variables: $A_{\text{ART: NNRTI}}$, $A_{\text{ART: bPI}}$, and $A_{\text{ART: DTG}}$. The residual is computed as 
\begin{align*}
    \tilde{A}_{\text{ART}: d} = A_{\text{ART:} d} - \hat{E}[A_{\text{ART:} d}| \bm{X}] = A_{\text{ART:} d} - \hat{p}[A_{\text{ART:} d}| \bm{X}],
\end{align*}
where $\hat{p}[A_{\text{ART:} d}| \bm{X}]$ is the estimated probability of ART type $d$ given covariates $\bm{X}$, for $d\in \{\text{NNRTI, bPI, DTG}\}$. This approach also applies to the treatment interaction $A^\circ_{\text{ART*TDF}}$, as detailed in Appendix Section 3. 

Importantly, we ensure that the coefficients in the residual regression can be interpreted as effects relative to the reference levels. Since the NNRTI serves as a reference level for the three-category ART treatment,  the residual $\tilde{A}_{\text{ART: NNRTI}}$ is omitted from the residual regression model. The same approach applies to the interaction term:  we omit the terms containing either the level of NNRTI or the level of ``no TDF use''. Therefore, the final residual regression is specified as
\begin{align}
\tilde{Y}_{\text{eGFR}} & = \theta_1 \tilde{A}_{\text{ART: bPI}} + \theta_2 \tilde{A}_{\text{ART: DTG}} + \theta_3 \tilde{A}_{\text{TDF}} + \theta_4 \tilde{A}_{\text{HTN}} \\
& \qquad + \theta_5 \tilde{A}^\circ_{\text{ART: bPI * TDF}} + \theta_6 \tilde{A}^\circ_{\text{ART: DTG * TDF}} \nonumber.
\end{align}
Here, each coefficient represents the effect of a treatment or interaction term on the outcome $Y_{\text{eGFR}}$. Specifically, $\theta_1$ and $\theta_2$ are the marginal effects of ARTs switching from NNRTI- to bPI- or DTG-based, respectively. $\theta_3$ and $\theta_4$ are the marginal effects of TDF and anti-HTN medications, respectively. The interaction terms $\theta_5$ and $\theta_6$ quantify how the effect of TDF changes when combined with bPI- or DTG-based ART, respectively. Since interaction effect is conditional on the main effect terms, all these coefficients will be interpreted jointly.

We use the boosting trees as the learner for both treatment and outcome models to account for complex confounding relationships. We use a 5-fold cross-fitting in the algorithm. To reduce dependence on a single data partition in cross-fitting, we repeat the algorithm 100 times and report the median estimate and the median standard error. 95\% CI are computed based on the median standard error.
\subsection{Results}

Table \ref{tab: application est coefs} summaries the estimated coefficients for main treatments and their interactions using our proposed model for multiple treatments. These estimated coefficients should be interpreted jointly. Therefore, Table \ref{tab: application effect diffs} presents the effect differences relative to the reference level of the combination of three treatments ($A_{\text{ART}} = \text{NNRTI}$, $A_{\text{TDF}} = 0$, $A_{\text{HTN}} = 0$). Across all ART strata, bPI or DTG leads to lower eGFR than NNRTI, particularly for DTG, which consistently shows the largest negative effects. For example, among participants with $A_\text{TDF} = 0$ or $A_\text{HTN} = 0$, those on DTG-based ART have an estimated eGFR reduction of -20.3 units (95\% CI: -30.3, -10.4), compared to those on NNRTI-based ART.

\renewcommand{\arraystretch}{1.5}
\begin{table}[H]
    \centering
    \begin{tabular}{ l c cc}
        \hline
       Exposures & Estimated coefficient & SE & 95\% CI \\
   \hline
       $A_{\text{ART}}$ (reference level: NNRTI) & &  & \\
         \quad bPI & -11.1 & 1.9 & (-14.7, -7.5) \\
         \quad DTG       & -20.3  & 5.1 & (-30.0, -10.6)  \\ 
        $A_{\text{TDF}}$ & -2.9   & 1.2 &  (-5.2, -0.6) \\
         $A_{\text{HTN}}$ & -0.6  & 2.4  & (-5.2, 4.0) \\ 
         $A_{\text{ART*TDF}}$ (interaction term) &  &    &   \\ 
        \quad bPI $\times$ TDF & 1.3   & 2.5  &  (-3.5, 6.0) \\ 
        \quad DTG $\times$ TDF  & 6.2  & 5.2  &   (-3.7, 16.1)  \\ 
        \hline
    \end{tabular}
    \caption{Estimates of coefficients of multiple treatments on kidney function outcome eGFR. $A_{\text{ART}}$: a three-level categorical variable of ART: (1) Non-Nucleoside Reverse Transcriptase Inhibitor (NNRTI), (2) boosted Protease Inhibitor (bPI), or (3) Dolutegravir (DTG). $A_{\text{TDF}}$: a binary indicator of Tenofovir Disoproxil Fumarate (TDF) use. $A_{\text{HTN}}$: a binary indicator of  anti-hypertension (HTN) medications use. SE, Standard Error. CI, confidence interval.}
    \label{tab: application est coefs}
\end{table}

The effects of TDF on eGFR vary across ART regimens. For instance, among participants receiving NNRTI-based ART and not using anti-HTN medications, TDF use is associated with a modest but statistically significant reduction in eGFR (-2.9; 95\% CI: -5.3, -0.6) compared with no TDF use. Interestingly, under the bPI-based ART, combining TDF appears to mitigate the negative effect from bPI-based ART: while bPI-based ART alone shows a statistically significant reduction in eGFR (-11.1, 95\% CI: -14.8, -7.4), adding TDF together results in a smaller and statistically nonsignificant effect (-7.9, 95\% CI: -18.7, 3.0).

\begin{table}[H]
\centering
\begin{tabular}{c l c c c}
\toprule
 & & \makecell{$A_{\text{ART}} = \text{NNRTI}$ \\ (reference level)}  & \makecell{$A_{\text{ART}} = \text{bPI}$ \\ } & \makecell{$A_{\text{ART}} = \text{DTG}$\\}  \\
\midrule
\multirow{2}{*}{\makecell{ $A_{\text{TDF}} = 0$ \\ (reference level)}}
  & $A_{\text{HTN}} = 0$ & \makecell{ 0 } & \makecell{-11.1 \\ (-14.8, -7.4)} & \makecell{-20.3 \\ (-30.3, -10.4)}  \\ \addlinespace[0.5 em]
  & $A_{\text{HTN}} = 1$ & \makecell{-0.6 \\ (-5.3, 4.0)} & \makecell{-11.7 \\ (-17.8, -5.6)} & \makecell{-21.0 \\ (-32.0, -9.9)} \\ \addlinespace[0.5 em]
\midrule
\multirow{2}{*}{$A_{\text{TDF}} = 1$ } 
  & $A_{\text{HTN}} = 0$ & \makecell{-2.9 \\ (-5.3, -0.6)} & \makecell{-7.9 \\ (-18.7, 3.0)} & \makecell{-21.9 \\ (-32.8, -11.1)} \\ \addlinespace[0.5 em]
  & $A_{\text{HTN}} = 1$ & \makecell{-3.6 \\ (-8.9, 1.8)} & \makecell{-8.5 \\ (-20.3, 3.3)} & \makecell{-22.6 \\ (-34.5, -10.6)} \\
\bottomrule
\end{tabular}
 \caption{Estimated effect differences relative to the reference group ($A_{\text{ART}} = \text{NNRTI}$, $A_{\text{TDF}} = 0$, $A_{\text{HTN}} = 0$). Each cell reports the contrast between the potential outcome under the treatment combination of the row and column versus that under the reference level. Values are presented as point estimates with 95\% confidence intervals in parentheses.}
    \label{tab: application effect diffs}
\end{table}

\section{Discussion}

Under the strong unconfoundedness assumption, the requirement for joint independence of potential outcomes across all treatments can be relaxed in many settings. This assumption helps to identify higher order interaction effects (e.g., three-way or higher interactions). However, if the true model only involves pairwise interactions, it suffices to assume pairwise conditional independence of potential outcomes across pairwise combinations of treatments.

The DML partial linear model implicitly assumes additive relationships between the multiple treatments and the outcome. This structure is not overly restrictive as it is close to traditional multi-way ANOVA, where treatments and their interactions are modeled linearly. Like ANOVA, the partial linear model distinguishes between main effects and interaction effects of treatments, allowing interaction effect to be interpreted conditional on the main treatments. However, DML further extends this structure by using machine learning to learn complex confounding models, improving the robustness of the estimates for treatment effects.

In sum, we describe (1) the DML partial linear model to estimate multiple treatments and their interactions and (2) the DML interactive to estimate the ATE between two categories of multi-valued regimens. We show that the score function of these models satisfy the Neyman orthogonality condition and thereby derive the variance estimator. The Nigeria HIV application shows that the proposed DML methods provide a useful tool for robust estimates of multiple treatments leveraging the strengths of machine learning.

%%%%%%%%%%%%%%%%%%%%%%%%%%%%%%%%%%%%%%%%%%%%%%
%% Appendix---Please move all appendices to %%
%% a Supplementary file.                    %%
%%%%%%%%%%%%%%%%%%%%%%%%%%%%%%%%%%%%%%%%%%%%%%
%% Support information, if any,             %%
%% should be provided in the                %%
%% Acknowledgements section.                %%
%%%%%%%%%%%%%%%%%%%%%%%%%%%%%%%%%%%%%%%%%%%%%%

\section*{Acknowledgments}
This study is funded in part by U01DK112271 and R01DK127912 (Nigeria HIV study), R01AI093234 (B.E.S), and R37AI093234 (B.E.S and C.W.W). The findings and conclusions are those of the authors and do not necessarily represent the official position of the National Institutes of Health, the Department of Health and Human Services or the government of the United States of America. The authors would like to thank the participants from Nigeria HIV study for their involvement.

\section*{Supplementary Material}
The Appendix includes the additional proofs, simulation details, and application details. The codes of this project are available at https://github.com/qingyan16.

\clearpage
\section*{Appendix}
\setcounter{section}{0}
\setcounter{subsection}{0}
\setcounter{equation}{0}
\setcounter{table}{0}
\setcounter{figure}{0}
\setcounter{assumption}{0}
\renewcommand{\thesection}{A.\arabic{section}}
\renewcommand{\thesubsection}{A.\arabic{section}.\arabic{subsection}}
\renewcommand{\theequation}{A.\arabic{equation}}
\renewcommand{\thetable}{A.\arabic{table}}
\renewcommand{\thefigure}{A.\arabic{figure}}
\renewcommand{\theassumption}{A.\arabic{assumption}}
\makeatletter
\renewcommand*{\theHsection}{appendix.\arabic{section}}
\renewcommand*{\theHsubsection}{appendix.\arabic{section}.\arabic{subsection}}
\renewcommand*{\theHequation}{appendix.\arabic{equation}}
\renewcommand*{\theHtable}{appendix.\arabic{table}}
\renewcommand*{\theHfigure}{appendix.\arabic{figure}}
\makeatother
\section{Theoretical details}

\subsection{Proof for Theorem 1}

\begin{proof}
The following equations  show that under the strong unconfoundedness assumption, the expected potential outcome \( E[Y_i^{(a_{i1},a_{i2},\dots,a_{iD})}] \) can be expressed using observed data.

\begin{align*}
    E[Y_i^{(a_{i1},a_{i2},\dots,a_{iD})}]
    &= E \left[ E[Y_i^{(a_{i1},a_{i2},\dots,a_{iD})} \mid \bm{X}_i] \right] \\
    & =  E \left[ E[Y_i^{(a_{i1},a_{i2},\dots,a_{iD})}\mid A_{i1} = a_{i1}, A_{i2} = a_{i2}, \dots, A_{iD} = a_{iD}, \bm{X}_i] \right] \\
    & = E \left[ E[Y_i \mid A_{i1} = a_{i1}, A_{i2} = a_{i2}, \dots, A_{iD} = a_{iD}, \bm{X}_i] \right].
\end{align*}

The first equality uses the law of iterated expectations. The second and the third equality use the strong unconfoundedness assumption.
\end{proof}

\subsection{Proof for Lemma 1} 

\begin{proof}

To verify Neyman orthogonality \citep{neyman1959optimal} for the score function
\[
    \psi(\bm{W}; \bm{\Theta}, \bm{\eta}) = \begin{bmatrix} \bm{A} -\bm{m}(\bm{X})\\ \bm{A}^\circ - \bm{m}^\circ(\bm{X}) \end{bmatrix} \left( Y - l(\bm{X}) -  \begin{bmatrix} \bm{A} -\bm{m}(\bm{X})\\ \bm{A}^\circ - \bm{m}^\circ(\bm{X}) \end{bmatrix}^\top \begin{bmatrix}
        \bm{\theta} \\
        \bm{\theta}^{\circ}
    \end{bmatrix} \right),
\]
we proceed by perturbing the nuisance parameters \(\eta = (\bm{m}, \bm{m}^\circ, l)\) along paths \(\bm{m} \to \bm{m} + t\bm{\delta}_m\), \(\bm{m}^\circ \to \bm{m}^\circ + t\bm{\delta}_{m^\circ}\), and \(l \to l + t\delta_l\), where \(\bm{\delta}_m, \bm{\delta}_{m^\circ}, \delta_l\) are arbitrary errors/perturbations. For simplicity, we omit $\bm{X}$ when it is not needed. The perturbed score is:
\[
\psi(t) = \begin{bmatrix} \bm{A} - \bm{m} - t\bm{\delta}_m \\ \bm{A}^\circ - \bm{m}^\circ - t\bm{\delta}_{m^\circ} \end{bmatrix} \cdot \left[ Y - l - t\delta_l - \begin{bmatrix} \bm{A} - \bm{m} - t\bm{\delta}_m \\ \bm{A}^\circ - \bm{m}^\circ - t\bm{\delta}_{m^\circ} \end{bmatrix}^T \begin{bmatrix} \bm{\theta} \\ \bm{\theta}^\circ \end{bmatrix} \right].
\]
We show that this score function satisfies the Neyman orthogonality condition in two steps: 
\begin{enumerate}
    \item  We calculate the derivative of this score function w.r.t $t$.
    \item We show that the expectation of this derivative is zero when $t = 0$. 
\end{enumerate}

First, let \(\varepsilon = Y - l - (\bm{A} - \bm{m})^\top \bm{\theta} - ( \bm{A^\circ} - \bm{m}^\circ)^\top \bm{\theta}^\circ\) such that,  w.r.t the $t$, 
\begin{align}
    \varepsilon(t) = \varepsilon + t\left(-\delta_l + \bm{\delta}_m^{\top} \bm{\theta} + \bm{\delta}_{m^\circ}^{\top}\bm{\theta}^\circ\right).
\end{align}
Then, the perturbed score can be expressed as
\[
\psi(t) = \begin{bmatrix} (\bm{A} - \bm{m}) - t\bm{\delta}_m \\ (\bm{A}^\circ - \bm{m}^\circ) - t\bm{\delta}_{m^\circ} \end{bmatrix} \cdot \left[ \varepsilon + t\left(-\delta_l + \bm{\delta}_m^{\top} \bm{\theta} + \bm{\delta}_{m^\circ}^{\top}\bm{\theta}^\circ \right) \right].
\]
Using the product rule, the derivative at \(t = 0\) is:
\begin{align}
    \label{eq: PLR score derivative}
    \frac{d\psi}{dt}\bigg|_{t=0} = \underbrace{\begin{bmatrix} -\bm{\delta}_m \\ -\bm{\delta}_{m^\circ} \end{bmatrix} \varepsilon}_{\text{Term 1}} + \underbrace{\begin{bmatrix} \bm{A} - \bm{m} \\ \bm{A}^\circ - \bm{m}^\circ \end{bmatrix} \left(-\delta_l +\bm{\delta}_m^{\top} \bm{\theta} + \bm{\delta}_{m^\circ}^{\top}\bm{\theta}^\circ \right)}_{\text{Term 2}}.
\end{align}

Second, we compute the expectation of the equation \eqref{eq: PLR score derivative}. For the term 1, by construction, \(\varepsilon\) satisfies \(E[\varepsilon \mid \bm{X}] = 0\); the perturbations $\bm{\delta}_m$ and $\bm{\delta}_{m^\circ}$ are functions of $\bm{X}$. Thus,
\begin{align*}
 E[\bm{\delta}_m \varepsilon] & =  E[\bm{\delta}_m(\bm{X})\varepsilon] \\
 & = E\left[ E[\bm{\delta}_m(\bm{X}) \varepsilon \mid \bm{X}]\right] \\
 & = E\left[ \bm{\delta}_m(\bm{X}) E[\varepsilon \mid \bm{X}]\right] \\
 & = \bm{0} ,  
\end{align*}
and this also applies for \(\bm{\delta}_{m^\circ}\). 
So the term 1 
\[
E\left[ \begin{bmatrix} -\bm{\delta}_m \\ -\bm{\delta}_{m^\circ} \end{bmatrix} \varepsilon \right] = \bm{0}.
\]
For the term 2, the residuals \(\bm{A} -\bm{m}(\bm{X})\) and \(\bm{A}^\circ - \bm{m}^\circ(\bm{X})\) satisfy \(E[\bm{A} -\bm{m}(\bm{X}) \mid \bm{X}] = \bm{0} \) and \(E[\bm{A}^\circ - \bm{m}^\circ(\bm{X}) \mid \bm{X}] = \bm{0} \); all perturbations (\(\delta_l, \bm{\delta}_m, \bm{\delta}_{m^\circ}\)) are functions of \( \bm{X}\). Thus,  for term $\delta_l$,
\[
E\left[ (\bm{A} -\bm{m}(\bm{X})) \cdot \delta_l(\bm{X}) \right] = E\left[ \delta_l(\bm{X}) \cdot E[\bm{A} -\bm{m}(\bm{X}) \mid \bm{X}] \right] = \bm{0}, 
\]
and this also applies for $\bm{\delta}_m$ and  $\bm{\delta}_{m^\circ}$. So the term 2
\begin{align*}
    E\left[ \begin{bmatrix} \bm{A} - \bm{m} \\ \bm{A^\circ} - \bm{m}^\circ \end{bmatrix} \left(-\delta_l + \bm{\delta}_m \bm{\theta} + \bm{\delta}_{m^\circ} \bm{\theta}^\circ\right) \right] = \bm{0}.
\end{align*}

In conclusion,  both terms in Equation \eqref{eq: PLR score derivative} have zero expectation. Since differentiation and expectation can be interchanged under Assumption A.1 (Appendix section 1.3), the zero expectation of the Equation  \eqref{eq: PLR score derivative} confirms the Neyman orthogonality condition:
\begin{align*}
    \partial_t E[\psi(\bm{W}; \bm{\Theta}_0, \bm{\eta}_0 + t\bm{\Delta})]\big|_{t=0} = \bm{0}.
\end{align*}

\end{proof}

\subsection{Proof for proposition 1}

We introduce Assumption A.1, adapted from the Assumption 4.1 in \citet{chernozhukov2018double}, for the proof of Proposition 1. 

\begin{assumption}{}
Let $\mathcal{P}$ be the collection of probability laws $P$ for $\bm{W} = (Y, \bm{A}, \bm{A}^\circ, \bm{X})$ such that 
\begin{itemize}
    \item[(a)] The treatment model (2) and the outcome model (3) in the main text holds;
    \item[(b)]  $\|Y\|_{P,q} + \| \bm{A} \|_{P,q}  + \|\bm{A}^\circ\|_{P,q}\leq C$;
    \item[(c)] $\|  \bm{\varepsilon_A} \varepsilon_Y \|_{P,2} \geq c^2$ and $E_P[\varepsilon_Y^2] \geq c$;
    \item[(d)]  $\|E_P[\varepsilon_Y^2 \mid \bm{X}]\|_{P,\infty} \leq C$ and  $ \max_{d}\|E_P[{\varepsilon_{A_d}}^2 \mid \bm{X}]\|_{P,\infty} \leq C$ ;
    \item[(e)] In the cross-fitting, given a random subset $I$ of $[N]$ of size $n = N/K$, the nuisance parameter estimator $\hat{\bm{\eta}}_0 = \hat{\bm{\eta}}_0((\bm{W}_i)_{i \in I^c})$ obeys the following conditions for all $n \geq 1$. With $P$-probability no less than $1 - \Delta_N$, $\|\hat{\bm{\eta}}_0 - \bm{\eta}_0\|_{P,q} \leq C$, $\|\hat{\bm{\eta}}_0 - \bm{\eta}_0\|_{P,2} \leq \delta_N$,  where $\hat{\bm{\eta}}_0 = (\hat{l}_0, \hat{\bm{m}}_0, \hat{\bm{m}}_0^\circ)$. For the score function (11) in the main text,
    $$
   \left \| \begin{bmatrix}
       \hat{\bm{m}}_{0} \\ 
       \hat{\bm{m}}_{0}^\circ
   \end{bmatrix}  - \begin{bmatrix}
       \bm{m}_{0} \\ 
       \bm{m}_{0}^\circ
   \end{bmatrix}   \right \|_{P,2} \cdot   \left( \left \| \begin{bmatrix}
       \hat{\bm{m}}_{0} \\ 
       \hat{\bm{m}}_{0}^\circ
   \end{bmatrix}  - \begin{bmatrix}
       \bm{m}_{0} \\ 
       \bm{m}_{0}^\circ
   \end{bmatrix}   \right\|_{P,2} + \|\hat{l}_0 - l_0\|_{P,2} \right) \leq \delta_N N^{-1/2}.
    $$
\end{itemize}
Here, $\| \cdot \|_{P,q}$ denotes the $L^{q}(P )$ norm; $(\delta_N)_{n=1}^{\infty}$ and $(\Delta_N)_{n=1}^{\infty}$ are the sequences of positive constants approaching 0 such that $\delta_N \geq N^{-1/2}$; $c$, $C$ and $q$ are fixed strictly positive constants such that $q > 4$; let $K \geq 2$ be a fixed integer; For simplicity, assume that $N/K$ is an integer.
\end{assumption}

The Assumption A.1 is closely aligned with the Assumption 4.1 in \citet{chernozhukov2018double} with two differences.  First, since we consider  a vector of $\bm{\varepsilon_A}$, we require that each component $\varepsilon_{A_d}$ is bounded (Assumption A.1 (d)). Second, we introduce the interaction term $\bm{A}^\circ$ and their model $\bm{m}_{0}^\circ$ in Assumption A.1 (b) and (e). However, if one  concatenate $\bm{A}$ and $\bm{A}^\circ$ into the same vector (and similarly $\bm{m}_{0}$ and $\bm{m}_{0}^\circ$),  (b) and (e) would be equivalent to the assumption 4.1 (b) and (e) in \citet{chernozhukov2018double}.  Now we sketch the proof for the Proposition 1.

\begin{proof}

Step 1, we begin by introducing a key result from Theorem 3.1 in \citet{chernozhukov2018double}. Consider the case of the linear   Neyman orthogonality score $\psi(W; \theta_0, \eta_0) = \psi_a(W, \eta_0)\theta + \psi_b(W, \eta_0)$, this theorem establishes the asymptotic normality of the DML estimator:
\begin{align*}
    \sqrt{N} (\hat{\theta} - \theta_0 ) \to N(0, \sigma^2),
\end{align*}
where  the approximate variance $\sigma^2$ is
$$
\sigma^2 := J_0^{-1} E [ \psi(W; \theta_0, \eta_0)\psi(W; \theta_0, \eta_0)^\top ] (J_0 ^{-1})^{\top}, 
$$
and the $J_{0}$ operator is
$$
J_{0} = E[\psi_a(W, \eta_0)].
$$

Step 2, we revisit the score function of the DML partial linear model for multiple treatments and their interactions:
\begin{align*}
    \psi(\bm{W}; \bm{\Theta}, \bm{\eta}) &= \begin{bmatrix} \bm{A} -\bm{m}(\bm{X})\\ \bm{A}^\circ - \bm{m}^\circ(\bm{X}) \end{bmatrix} \left( Y - l(\bm{X}) -  \begin{bmatrix} \bm{A} -\bm{m}(\bm{X})\\ \bm{A}^\circ - \bm{m}^\circ(\bm{X}) \end{bmatrix}^\top \begin{bmatrix}
        \bm{\theta} \\
        \bm{\theta}^{\circ}
    \end{bmatrix} \right) \\
     & = - \begin{bmatrix} \bm{A} -\bm{m}(\bm{X})\\ \bm{A}^\circ - \bm{m}^\circ(\bm{X}) \end{bmatrix} \begin{bmatrix} \bm{A} -\bm{m}(\bm{X})\\ \bm{A}^\circ - \bm{m}^\circ(\bm{X}) \end{bmatrix}^\top \begin{bmatrix}
        \bm{\theta} \\
        \bm{\theta}^{\circ}
    \end{bmatrix} + \begin{bmatrix} \bm{A} -\bm{m}(\bm{X})\\ \bm{A}^\circ - \bm{m}^\circ(\bm{X}) \end{bmatrix}  (Y - l(\bm{X})).
\end{align*}
Since the score function is linear in $\theta$, following main text Equation (9), it can be decomposed into:
\begin{align*}
    \psi_a(\bm{W}, \bm{\eta}) &= - \begin{bmatrix} \bm{A} -\bm{m}(\bm{X})\\ \bm{A}^\circ - \bm{m}^\circ(\bm{X}) \end{bmatrix} \begin{bmatrix} \bm{A} -\bm{m}(\bm{X})\\ \bm{A}^\circ - \bm{m}^\circ(\bm{X}) \end{bmatrix}^\top, \\ 
    \psi_b(\bm{W}, \bm{\eta}) &=  \begin{bmatrix} \bm{A} -\bm{m}(\bm{X})\\ \bm{A}^\circ - \bm{m}^\circ(\bm{X}) \end{bmatrix}  (Y - l(\bm{X})) .
\end{align*}

Step 3, we apply Theorem 3.1 in \citet{chernozhukov2018double} to our score function, which leads to an estimator that satisfies:
\[
\sqrt{n} \left( \begin{bmatrix}
    \hat{\bm{\theta}} \\
    \hat{\bm{\theta}}^\circ
\end{bmatrix}  - \begin{bmatrix}
    \bm{\theta}_0 \\
    \bm{\theta}^\circ_0
\end{bmatrix} \right) \overset{d}{\to} N\left( \bm{0}, \, \bm{\Sigma} \right)
\]
with asymptotic variance-covariance matrix
\begin{align}
\label{eq: PLR variance}
    \bm{\Sigma} =  \bm{J}_0^{-1} E\left[\psi(\bm{W}; \bm{\Theta}_0, \bm{\eta}_0) \psi(\bm{W}; \bm{\Theta}_0, \bm{\eta}_0)^\top \right] (\bm{J}_0^{-1})^\top,
\end{align}
where 
\begin{align*}
       \bm{J}_0 = E[\psi_a(\bm{W}, \bm{\eta}_0)] 
    & = - E \left[ \begin{bmatrix} \bm{A} -\bm{m}(\bm{X})\\ \bm{A}^\circ - \bm{m}^\circ(\bm{X}) \end{bmatrix} \begin{bmatrix} \bm{A} -\bm{m}(\bm{X})\\ \bm{A}^\circ - \bm{m}^\circ(\bm{X}) \end{bmatrix}^\top \right] \\
    & =  -E \left[\begin{bmatrix}
        \bm{\varepsilon_A} \\
        \bm{\varepsilon_{A^\circ}}
    \end{bmatrix}  \begin{bmatrix}
        \bm{\varepsilon_A}^\top \; \bm{\varepsilon_{A^\circ}}^\top
    \end{bmatrix}   \right].
\end{align*}

Step 4, to conclude, applying Theorem 3.1 requires Assumptions 3.1 and 3.2 in \citet{chernozhukov2018double} to be satisfied. Assumption 3.1 requires the score \(\psi(\bm{W}; \bm{\Theta}_0, \bm{\eta}_0) \) satisfies the Neyman orthogonality condition, which is verified in Lemma 1. Assumption 3.2 requires the quality of the nuisance estimators \(\hat{\bm{m}}\), \(\hat{\bm{m}}^\circ\), and \(\hat{l}\): they  converge to the true values at a rate of \(o_p(n^{-1/4})\). Assumption A.1 verifies the Assumption 3.2 following the proof in Theorem 4.1 in \citep{chernozhukov2018double}, and the rate of \(o_p(n^{-1/4})\) is achievable via most machine learning methods under structural assumptions.
\end{proof}

\subsection{Proof for Lemma 2}
\begin{proof}
    This proof shows that the score function (13) for DML interactive model for multi-valued regimens satistifies the Neyman orthogonality.
    
    For simplicity, we omit $\bm{X}$ throughout most of the text and use $g_b$ to denote $g(r=b, \bm{X})$ in the main text. We proceed by perturbing the nuisance parameters: $m_b \to m_b + t\delta_{m_b}$, $m_c \to m_c + t\delta_{m_c}$, $g_b \to g_b + t\delta_{g_b}$, and $g_c \to g_c + t\delta_{g_c}$, where $\delta_{m_b},\delta_{g_b}, \delta_{g_b}, \delta_{g_c}$ are arbitrary errors/perturbations. 

The perturbed score becomes:
\[
\psi(t) = \left[ g_b - g_c + t(\delta_{g_b} - \delta_{g_c}) \right] + \frac{\mathbbm{1}_{R=b}\left(Y - g_b - t\delta_{g_b}\right)}{m_b + t\delta_{m_b}} - \frac{\mathbbm{1}_{R=c}\left(Y - g_c - t\delta_{g_c}\right)}{m_c + t\delta_{m_c}} - \theta.
\]
Next, we take the derivative of the score function w.r.t $t$:

\begin{align*}
     \frac{d\psi}{dt} =  (\delta_{g_b} - \delta_{g_c})  -\frac{\mathbbm{1}_{R = b} \left[ \delta_{g_b} m_b + \delta_{m_b} (Y - g_b) \right]}{(m_b + t \delta_{m_b})^2}  + \frac{\mathbbm{1}_{R = c} \left[ \delta_{g_c} m_c + \delta_{m_c} (Y - g_c) \right]}{(m_c + t \delta_{m_c})^2},
\end{align*}
and hence the derivative at \(t = 0\) is
\begin{align}
\label{eq: IRM score derivative}
    \frac{d\psi}{dt}\bigg|_{t=0} & = (\delta_{g_b} - \delta_{g_c}) - \mathbbm{1}_{R=b}\left( \frac{\delta_{g_b}}{m_b} + \frac{(Y - g_b)\delta_{m_b}}{m_b^2} \right) + \mathbbm{1}_{R=c}\left( \frac{\delta_{g_c}}{m_c} + \frac{(Y - g_c)\delta_{m_c}}{m_c^2} \right). \\
& =   \underbrace{(\delta_{g_b} - \delta_{g_c})}_{\text{Term 1}} \underbrace{-\mathbbm{1}_{R=b} \frac{\delta_{g_b}}{m_b} +  \mathbbm{1}_{R=c} \frac{\delta_{g_c}}{m_c}}_{\text{Term 2}}  \underbrace{-\mathbbm{1}_{R=b} \frac{(Y - g_b)\delta_{m_b}}{m_b^2} +  \mathbbm{1}_{R=c} \frac{(Y - g_c)\delta_{m_c}}{m_c^2}}_{\text{Term 3}} \nonumber
\end{align}

After taking expectations of the Equation \eqref{eq: IRM score derivative}, the term 1 \(E[\delta_{g_b} - \delta_{g_c}]\) cancels with term 2, as will be shown later.

For term 2,
\begin{align*}
    E\left[-\mathbbm{1}_{R=b} \frac{\delta_{g_b}}{m_b} +  \mathbbm{1}_{R=c} \frac{\delta_{g_c}}{m_c} \right] & = - E\left[E\left[ \mathbbm{1}_{R=b} \mid \boldsymbol{X}\right] \frac{\delta_{g_b}}{m_b}\right] + E\left[E\left[ \mathbbm{1}_{R=c} \mid \boldsymbol{X}\right] \frac{\delta_{g_c}}{m_c}\right] \\
   &  = - E[\delta_{g_b}] + E[\delta_{g_c}].
\end{align*}
The second equation follows from  $E[ \mathbbm{1}_{R=b} \mid \boldsymbol{X} ] = P(R=b| \bm{X}) = m_b$ (same for $m_c$).

For term 3,
\begin{align*}
    & E\left[-\mathbbm{1}_{R=b} \frac{(Y - g_b)\delta_{m_b}}{m_b^2} +  \mathbbm{1}_{R=c} \frac{(Y - g_c)\delta_{m_c}}{m_c^2} \right] \\
    & = - E\left[ \frac{\delta_{m_b}}{m^2_b} \cdot \underbrace{E\left[\mathbbm{1}_{R=b}(Y - g_b) \mid \boldsymbol{X}\right]}_{=0}\right] + E\left[\frac{\delta_{m_c}}{m^2_c} \cdot \underbrace{E\left[\mathbbm{1}_{R=c}(Y - g_c) \mid \boldsymbol{X}\right]}_{=0}\right] \\
    & = 0.
\end{align*}

Therefore, combining all three terms, the total expectation of Equation \eqref{eq: IRM score derivative} is 
\[
\mathbb{E}\left[\frac{d\psi}{dt}\bigg|_{t=0}\right] = \mathbb{E}[\delta_{g_b} - \delta_{g_c}] - \mathbb{E}[\delta_{g_b}] + \mathbb{E}[\delta_{g_c}] = 0.
\]
This zero expectation confirms the Neyman orthogonality condition.

\end{proof}

\subsection{Proof for proposition 2}
The assumption A.2, adapted from on the Assumption 6.1 in \citet{chernozhukov2018double}, is required for the proofs of Proposition (2).

\begin{assumption}
    For all probability laws \( P \in \mathcal{P} \) of the triple \((Y, R, \bm{X})\) and for the target regimens $b$ and $c$, the following conditions hold:
\begin{itemize}
        \item[(a)] The treatment model (5) and the outcome model (6) in the main text holds;
    \item[(b)] \( \|Y\|_{P,q} \leq C \);
    \item[(c)] Positivity assumption in the main text holds.
    \item[(d)] \( \|\varepsilon_Y\|_{P,2} \geq c \);
    \item[(e)] \( \| \mathbb{E}_P[\varepsilon_Y^2 \mid \bm{X}] \|_{P,\infty} \leq C \);
    \item[(f)] Given a random subset \( I \) of \([N]\) of size \( n = N/K \), the nuisance parameter estimator \( \hat{\eta} = \hat{\eta}((W_i)_{i \in I^c}) \) obeys the following conditions. With \( P \)-probability no less than \( 1 - \xi_N \), \( \|\hat{\eta} - \eta_0\|_{P,q} \leq C \),  \( \|\hat{\eta} - \eta_0\|_{P,2} \leq \delta_N \),  \( \|\hat{m}_d - 1/2\|_{P,\infty} \leq 1/2 - \varepsilon \) for $d\in{b, c}$, and \( \|\hat{m}_d - m_{d}\|_{P,2} \times \|\hat{g}_d - g_{d}\|_{P,2} \leq \delta_N N^{-1/2} \) for  $d \in {b, c}$.
\end{itemize}
Here, \((\delta_N)_{n=1}^{\infty}\) and \((\xi_N)_{n=1}^{\infty}\) are sequences of positive constants approaching 0. \(c\), \(\varepsilon\), \(C\), and \(q\) are fixed strictly positive constants such that \(q > 2\), and let \(K \geq 2\) be a fixed integer. For simplicity, assume that \(N/K\) is an integer.
\end{assumption}

Our score function closely resembles the one used in \citet{chernozhukov2018double}, with the key difference being the use of generalized propensity scores $m_b$ and $m_c$ in place of the binary propensity scores $m$ and $1 - m $. Given this structural similarity, the proofs of our Proposition 2 follow directly from Theorem 5.1 in \citet{chernozhukov2018double} with minimal modifications. For brevity, we do not reproduce the proofs here.

\newpage
\section{Additional simulation details}
\subsection{DML partial linear model for multiple treatments and their interactions}
The covariates are generated as follows. $X_1,...,X_5$ are all generate with normal distribution $N(0,1)$. $X_6, ..., X_{10}$ are generated as binomial distribution as 
\begin{align*}
    X_k \sim Binom(\pi_k),
\end{align*}
where
\begin{align*}
    \pi_6 = 0.1, \pi_7 = 0.3, \pi_8 = 0.5, \pi_9 = 0.7, \pi_{10} = 0.9.
\end{align*}
The detailed data generation for treatment model and outcome model is specified in the main text.

\iffalse
As mentioned in the main text, we generate two treatments:
\begin{itemize}
    \item a binary treatment generated as $A_1 \sim \text{Bernoulli} (\pi_1)$, where  $\text{logit}(\pi_1) = m_1(\bm{X})$. 
    \item a continuous treatment $A_2$ generated by $A_2 = m_2(\bm{X}) + \varepsilon_{A_2}$, where $ \varepsilon_{A_2} \sim N(0, 1)$.
\end{itemize}
The functions \( m_1(\bm{X}) \) and \( m_2(\bm{X}) \) specify nonlinear relationships between covariates and treatment assignment mechanisms:
\begin{align*}
    m_1(\bm{X}) & = 0.8 X_1 X_2 + 0.4 X_2^2 + 0.4 X_3 + 0.7 X_4 + 0.3 X_6 + 0.9 X_7 X_8 - 1.3 X_9. \\
    m_2(\mathbf{X}) & = \frac{1}{1 + \exp(X_1)} - \frac{1}{1 + \exp(X_2)} + 0.5 X_3 + 0.25 \left( \mathbbm{1}(X_5 > 0) - \mathbbm{1}(X_6 > 0) \right) + 0.1 \left( X_7 + X_9 X_{10} \right).
\end{align*}

The outcome model is specified by
\begin{align*}
    Y= \theta_{1}  A_1 + \theta_{2}   A_2+ \theta_{3}  A_1  A_2 + g(\bm{X}) + \varepsilon,
\end{align*}
where 
\begin{align*}
    g(\bm{X}) =  \bm{X}^L \bm{\beta}^L_d + \bm{X}^{NL} \bm{\beta}^{NL}_d & =  -2 \mathbbm{1}(X_1 < 0) +  2\mathbbm{1}(X_1 \geq 0)  -  \mathbbm{1}(X_2 < 1) +  \mathbbm{1}(X_2 \geq 1) \\
& \qquad + 2X_3 + 2X_5 + X_6 + X_7 - 2X_9 - 0.5X_{10} 
\end{align*}

\fi

The machine learning models are specified as follows. Note $A_1$ is a binary treatment while $A_2$, $A_1\times A_2$, and $Y$ are continuous; we use classification model for $A_1$ and regression models for $A_2$, $A_1\times A_2$, and $Y$. We tune our machine learning models using grid search. Random forests were implemented using the R package \texttt{ranger}, with the minimum node size set to 1 (classification) or 5 (regression), and all other parameters kept at default values. Boosting trees were implemented using the R package \texttt{gbm}. For classification boosting trees, we set tree numbers to 100, shrinkage to 0.05, and minimum node size to 10, with other parameters at default. For regression boosting trees, we set tree numbers to 500, shrinkage to 0.01, interaction depth to 5, minimum node size to 1, with other parameters at default. Neural networks were implemented using the R package \texttt{nnet} (only one single hidden layer). For both classification and regression, we set  hidden neurons to 16, decay to 0.1, maximum iterations to 500, with other parameters at default. For the \texttt{nnet} The input covariate matrix $\bm{X}$ is processed with min-max normalization.

\subsection{DML interactive model for multi-value regimens}

The covariates are generated as follows. $X_1,...,X_5$ are all generate with normal distribution $N(0,1)$. $X_6, ..., X_{10}$ are generated as binomial distribution as 
\begin{align*}
    X_k \sim Binom(\pi_k),
\end{align*}
where
\begin{align*}
    \pi_6 = 0.1, \pi_7 = 0.3, \pi_8 = 0.5, \pi_9 = 0.7, \pi_{10} = 0.9.
\end{align*}

As mentioned in the main text, the regimens are generated from a multinomial distribution:
\begin{align*}
        R \mid X \sim \text{Multinomial} \left(N, m_1(\bm{X}), m_2(\bm{X}), m_3(\bm{X}) \right),
\end{align*}
where the probability of assigning treatment $a$ is determined by the softmax function:
\begin{align*}
    m_d(\bm{X}) = \frac{e^{l_d(\bm{X})}}{\sum_{d=1}^{3} e^{l_d(\bm{X})}}, \quad a = 1,2,3.
\end{align*}
We specify the log-linear predictor functions as follows:
\begin{align*}
    l_1(\bm{X}) &= 1, \\
    l_2(\bm{X}) &= 0.8 X_1 X_2 + 0.4 X_2^2 - 0.4 X_3 + 0.7 X_4 + 0.3 X_6 + 0.9 X_7 X_8 - 1.3 X_9, \\
    l_3(\bm{X}) &= -1.2 X_1 X_2 + 1.8X_3 + 2.5\mathbbm{1}(X_4>0) + 0.3 X_6 X_7 - 1.2 X_8 + 0.5 X_5 X_{10}.
\end{align*}
Here, $l_d(\bm{X})$ represents the linear predictor for treatment category $d$. The reference category is $R=1$, which is treated as the control regimen due to that $l_1(\bm{X}) = 1$. 

The outcome model of $Y$ is generated as 
\begin{align*}
   Y = g(R, \bm{X}) = 5 \cdot \mathbbm{1}_{R=2}+15\cdot \mathbbm{1}_{R=3} \cdot X_9 + \bm{X}^L \bm{\beta}^L_d + \bm{X}^{NL} \bm{\beta}^{NL}_d,
\end{align*}
where
\begin{align*}
\bm{X}^L \bm{\beta}^L_d + \bm{X}^{NL} \bm{\beta}^{NL}_d & =  -5 \mathbbm{1}(X_1 < 0) + 5 \mathbbm{1}(X_1 \geq 0)  - 8 \mathbbm{1}(X_2 < 1) + 8 \mathbbm{1}(X_2 \geq 1) \\
& \qquad + 2X_3 + 4X_5 + X_6 + 2X_7 + 4X_9 + 5X_{10} \\
& \qquad + 4X_3 X_4 + 6X_5 X_{10} + 6X_5^2 + 4X_9^2.
\end{align*}
The ATEs can be derived consequently:
\begin{align*}
    ATE_{21} & = E[Y^{(r = 2)}] - E[Y^{(r = 1)}] = 5, \\ 
    ATE_{31} & = E[Y^{(r = 3)}] - E[Y^{(r = 1)}] = 15 \cdot E[X_9] = 15 \cdot \pi_9 = 10.5, \\
    ATE_{32} & = E[Y^{(r = 3)}] - E[Y^{(r = 2)}] = 15 \cdot E[X_9] - 5 = 5.5. 
\end{align*}

For our proposed DML interactive model, we use GBM as the learners to model for both the outcome and treatment models. The outcome model is based on regression trees, while the treatment model uses classification trees to estimate the multivariate probabilities for multi-valued regimens.

For traditional causal inference methods, Inverse Propensity Score Weighting (IPW) and Propensity Score Matching (PSM), we use two type of learners: generalized linear models (with a logit link) and GBM. These learners are used solely for estimating the propensity scores and are not applied to outcome modeling. For a fair comparison in all cases, we adopt the default parameter settings of GBM. 

\newpage
\section{Additional application details}

\subsection{Summary table for patients' information}
\begin{table}[H]
    \renewcommand{\arraystretch}{1.3}
    \resizebox{\textwidth}{!}{%
    \begin{tabular}{lcccc}
        \toprule
        \textbf{Characteristic} &  \makecell{\textbf{Overall} \\ \textbf{(N = 2455)}} & \makecell{\textbf{NNRTI} \\ \textbf{(N = 1435)}} & \makecell{\textbf{Boosted-PI} \\ \textbf{(N = 431)}} & \makecell{\textbf{DTG} \\  \textbf{(N = 589)}} \\
        \midrule
        Age & 40 (34, 47) & 39 (34, 46) & 39 (34, 47) & 43 (36, 49) \\
        Gender (female)  & 1,712 (70\%) & 1,012 (71\%) & 297 (69\%) & 403 (68\%) \\
        Ethnicity & & & & \\
        \quad Hausa-Fulani & 1,767 (72\%) & 1,057 (74\%) & 292 (68\%) & 418 (71\%) \\
        \quad Igbo & 112 (4.6\%) & 58 (4.0\%) & 28 (6.5\%) & 26 (4.4\%) \\
        \quad Yoruba & 55 (2.2\%) & 25 (1.7\%) & 12 (2.8\%) & 18 (3.1\%) \\
        \quad Other & 521 (21\%) & 295 (21\%) & 99 (23\%) & 127 (22\%) \\
        Risk Alleles & & & & \\
        \quad 0 allele & 1,682 (69\%) & 981 (68\%) & 292 (68\%) & 409 (69\%) \\
        \quad 1 allele & 621 (25\%) & 364 (25\%) & 107 (25\%) & 150 (25\%) \\
        \quad 2 allele & 152 (6.2\%) & 90 (6.3\%) & 32 (7.4\%) & 30 (5.1\%) \\
        Smoking use & 122 (5.0\%) & 67 (4.7\%) & 18 (4.2\%) & 37 (6.3\%) \\
        \textbf{Tenofovir use} & 1,445 (59\%) & 577 (40\%) & 297 (69\%) & 571 (97\%) \\
        \textbf{Anti-hypertension medication use} & 279 (11\%) & 163 (11\%) & 40 (9.3\%) & 76 (13\%) \\
        Duration on ART, months & 9 (6, 12) & 9 (6, 12) & 11 (8, 13) & 8 (4, 11) \\
        CD4 count, cells/mm3 & 479 (319, 656) & 508 (353, 684) & 383 (251, 546) & 454 (293, 623) \\
         Recent Viral Load, <200 copies/ml & 105 (4.3\%) & 48 (3.3\%) & 44 (10\%) & 13 (2.2\%) \\
        Diabetes mellitus condition  & 50 (2.0\%) & 26 (1.8\%) & 11 (2.6\%) & 13 (2.2\%) \\

        Hypertension condition  & 361 (15\%) & 215 (15\%) & 48 (11\%) & 98 (17\%) \\
        Other comorbid conditions & 544 (22\%) & 308 (21\%) & 118 (27\%) & 118 (20\%) \\
        BMI & 23.2 (20.2, 26.9) & 23.1 (20.2, 26.7) & 23.0 (19.8, 26.7) & 23.6 (20.3, 27.5) \\
        Mean systolic BP, mm Hg & 110 (99, 123) & 111 (100, 125) & 107 (96, 119) & 109 (99, 122) \\
        Mean diastolic BP, mm Hg & 73 (66, 81) & 73 (67, 82) & 70 (63, 78) & 73 (67, 80) \\
        eGFR, ml/min per 1.73 m$^2$ & 104 (87, 120) & 111 (96, 125) & 99 (84, 116) & 90 (75, 105) \\
        \bottomrule
    \end{tabular}
    }
    \caption{Summary of demographic and clinical characteristics of the patients. Continuous variables are summarized using median (interquartile range), and categorical variables are summarized using n (\%). ART, antiretroviral therapy; BMI, body mass index; BP, blood pressure; uACR, urine albumin--creatinine ratio; eGFR, estimated glomerular filtration rate.}
\end{table}

This application uses data from \cite{wudil2021apolipoprotein}, a study that explored the relationship between \textit{apolipoprotein 1 (APOL1)} and kidney function among a cohort of HIV-positive Nigerian adults. The eligibility criteria for the participants includes: (i) HIV-positive, (ii) on ART for a minimum of 6 months, and (iii) between 18 and 70 years of age, as detailed in the study protocol \citet{aliyu2019optimal}. Table A.1 presents the demographic and clinical information measured in this study.

The summary table presents demographic and clinical characteristics of 2,455 participants with HIV, stratified by antiretroviral therapy (ART) regimen: NNRTI, Boosted-PI, and DTG. The variables \textit{Tenofovir use} and \textit{Anti-hypertension medication use} correspond to $A_{\text{TDF}}$ and $A_{\text{HTN}}$ as defined in the main text, respectively. There are 59\% of participants on Tenofovir, but less participants on anti-hypertension medication. The outcome is measured by estimated glomerular filtration rate (eGFR), an indicator for kidney function.

The median age overall was 40 years, with DTG users slightly older (median 43). Women made up the majority across all groups (70\% overall). Hausa-Fulani was the predominant ethnicity (72\%).  CD4 counts were highest among NNRTI users, while kidney function, as measured by eGFR, was slightly lower in the DTG group. Overall, comorbid conditions and risk allele distributions were similar across the groups, with some variations in metabolic and cardiovascular markers.

\subsection{Estimate the effect of six-level interaction term}

The interaction term $A^\circ_{\text{ART} * \text{TDF}}$ is formed by a producted of a three-level treatment $A_{\text{ART}}$ ((NNRTI, PI-boosted, DTG)) and a binary treatment $A_{\text{TDF}}$ (off TDF = 0, on TDF = 1). Therefore, $A^\circ_{\text{ART} * \text{TDF}}$ has six levels. We first apply dummy variable encoding to this interaction term $A^\circ_{\text{ART} * \text{TDF}}$. To illustrate, consider four hypothetical patients with different combinations of ART regimen and TDF usage, as shown in the Table A.2.

\begin{table}[H]
    \centering
    \begin{tabular}{|ccc|}
        \hline
        ID &  $A_{\text{ART}}$  & $A_{\text{TDF}}$ \\
        \hline
        1 &  \text{NNRTI} & 0 \\
        2 & \text{DTG} & 0  \\
        3 & \text{NNRTI} & 1  \\
        4 & \text{PI-boosted} & 1 \\
        \hline
    \end{tabular}
    \label{tab:patient_data}
    \caption{Illustrative data of four example patients with their usage of ART and TDF.}
\end{table}

We apply dummy variable encoding to this six-level categorical variable $A^\circ_{\text{ART} * \text{TDF}}$, resulting in six binary indicator columns---each representing a specific combination of ART and TDF. The resulting design matrix for these dummy variables is shown in Table A.3. For example, Patient 1 is on NNRTI and not on TDF, so only the column for ``$A_{\text{ART}}$ =  NNRTI and $A_{\text{TDF}}=0$'' is 1; all other columns are 0. Patient 4 is on PI-boosted ART and on TDF, so only the column for ``$A_{\text{ART}}$ = PI-boosted and $A_{\text{TDF}}=1$'' is 1.

\begin{table}[H]
    \centering
    \resizebox{\textwidth}{!}{%
    \begin{tabular}{|c|c|c|c|c|c|c|}
        \hline
        ID & \makecell{ $A_{\text{ART}}$ =  NNRTI \\ $A_{\text{TDF}}=0$} & \makecell{$A_{\text{ART}}$ = PI-boosted \\ $A_{\text{TDF}}=0$} & \makecell{$A_{\text{ART}}$ = DTG \\ $A_{\text{TDF}}=0$} & \makecell{$A_{\text{ART}}$= NNRTI \\ $A_{\text{TDF}}=1$} & \makecell{$A_{\text{ART}}$ = PI-boosted \\ $A_{\text{TDF}}=1$} & \makecell{$A_{\text{ART}}$= DTG \\ $A_{\text{TDF}}=1$} \\
        \hline
        1 & 1 & 0 & 0 & 0 & 0 & 0 \\
        2 & 0 & 0 & 1 & 0 & 0 & 0 \\
        3 & 0 & 0 & 0 & 1 & 0 & 0 \\ 
        4 & 0 & 0 & 0 & 0 & 1 & 0 \\ 
        \hline
    \end{tabular}
    \label{tab:treatment_groups}}
    \caption{Dummy variable encoding of the six-level interaction term $A^\circ_{\text{ART} * \text{TDF}}$ for four hypothetical patients.}
\end{table}

Next,  using the dummy variables in Table A.3 as the outcomes, we use a six-level classification model to estimate the probability for each category. Specifically, we use a gradient boosting machine as our model (R package \texttt{gbm}), with the minimum number of tree nodes set to 10 and all other parameters left at their defaults. For each column in Table A.3, the model estimates the probability that a patient falls into that category of interaction combination. By subtracting these predicted probabilities from the the dummy variables, we obtain six columns of residuals.

As we mentioned in the main text, if the binary dummy variable include the reference level for either $A_{\text{ART}}$ or $A_{\text{TDF}}$, they will not be included in the final residual regression. Since NNRTI serves as the reference level for $A_{\text{ART}}$ and "off TDF" serves as the reference level for $A_{\text{TDF}}$, the interaction term containing either $A_{\text{ART}} =$ NNRTI or $A_{\text{TDF}}=0$ should be omitted from the final residual regression. Therefore, only the residual terms of last two columns in Table A.3 are included in the residual regression, where we denote them as $\tilde{A}^\circ_{\text{ART: bPI * TDF}}$ and $\tilde{A}^\circ_{\text{ART: DTG * TDF}}$.

Finally, together with the terms of main effect, the residual regression is:
\begin{align*}
\tilde{Y}_{\text{eGFR}} & = \theta_1 \tilde{A}_{\text{ART: bPI}} + \theta_2 \tilde{A}_{\text{ART: DTG}} + \theta_3 \tilde{A}_{\text{TDF}} + \theta_4 \tilde{A}_{\text{HTN}} \\
& \qquad + \theta_5 \tilde{A}^\circ_{\text{ART: bPI * TDF}} + \theta_6 \tilde{A}^\circ_{\text{ART: DTG * TDF}}.
\end{align*}
This is also showed in main text Equation (15)

\subsection{Interpretation of the estimated effect}

Table 4 summarizes the estimated differences in $Y_{\text{eGFR}}$ across combinations of ART regimen, TDF use, and antihypertensive (HTN) medication, relative to the reference group $(A_{\text{ART}}=\text{NNRTI}, A_{\text{TDF}}=0, A_{\text{HTN}}=0)$. 
Each cell reports the contrast between the potential outcome under a given treatment combination and that under the reference level, with 95\% confidence intervals in parentheses.

These contrasts are directly derived from the estimated coefficients $\hat{\theta}_1$--$\hat{\theta}_6$ in the residual regression.  Since interaction effect is conditional on the main effect terms, all these coefficients will be interpreted jointly. For example, the effect of using both bPI-based ART and TDF (with no anti-HTN medication) is computed as the sum $\hat{\theta}_1 + \hat{\theta}_3 + \hat{\theta}_5$, corresponding to an estimated decrease of $-7.9$ units (95\% CI: $-18.7$, 3.0) relative to the reference. 
Similarly, DTG-based ART combined with TDF (with no anti-HTN medication) corresponds to $\hat{\theta}_2 + \hat{\theta}_3 + \hat{\theta}_6$, yielding a larger estimated decrease of $-21.9$ (95\% CI: $-32.8$, $-11.1$).

\end{document}